\documentclass[final,5p,times,twocolumn]{elsarticle}

\usepackage{amssymb}
\usepackage{import}
\usepackage{bm}
\usepackage[hidelinks]{hyperref}
\usepackage{optidef}
\usepackage{units}
\usepackage{times} %
\usepackage{amsmath} %
\usepackage{graphicx} %
\usepackage{multirow}
\usepackage{booktabs}
\usepackage{colortbl}
\usepackage{makecell}
\usepackage{algorithmicx}
\usepackage{algpseudocode}
\usepackage{pifont}
\usepackage{hhline}
\usepackage[export]{adjustbox}
\usepackage{stfloats}		%
\usepackage{accents}
\usepackage{color,soul}
\usepackage{xcolor}
\usepackage{lipsum}
\usepackage{pifont}
\usepackage{subcaption}

\RequirePackage{pgfplots}
\pgfplotsset{compat=1.12}
\graphicspath{{images/}}	%
\usepackage{lscape}

\definecolor{grey_manual}{RGB}{200,200,200}

\journal{arxiv.org}

\begin{document}

\begin{frontmatter}
\title{Energy Management of Hydrogen Hybrid Electric Vehicles - A Potential Study}

\author{David Theodor Machacek$^a$}
\ead{davidm@ethz.ch}
\author{Nazim Ozan Yazar$^a$}
\author{Thomas Huber$^b$}
\author{Christopher Harald Onder$^a$}

\affiliation{organization={Institute for Dynamic Systems and Control},%
            addressline={Sonneggstrasse 3}, 
            city={Zürich},
            postcode={8092}, 
            country={Switzerland}}
\affiliation{organization={Robert Bosch GmbH},%
            city={Stuttgard},
            postcode={70442}, 
            country={Germany}}

\begin{abstract}
The hydrogen combustion engine (H$_2$ICE) is known to be able to burn H$_2$ under ultra-lean conditions, while producing no CO$_2$ emissions and extremely low engine-out NO$_x^{\mathrm{eo}}$ emissions. This makes the H$_2$ICE a promising technology for two reasons: Firstly, mid-term goals such as net-zero CO$_2$ emissions of passenger cars in Europe by 2035 can be achieved. Secondly, immediate goals, as for instance the upcoming EURO 7 NO$_x$ limitations, can be reached more easily as extremely low engine-out NO$_x^{\mathrm{eo}}$ emissions facilitate the reduction of the overall tailpipe NO$_x^{\mathrm{tp}}$ emissions. However, one drawback of the H$_2$ICE is that for high engine loads, ultra-lean combustion is no longer possible, which partly nullifies its advantage over conventional combustion engines. 
In this work, the feasibility of achieving consistent reductions in NO$_x^{\mathrm{eo}}$ emissions through the implementation of electric hybridization of an H$_2$ICE-equipped passenger car (H$_2$-HEV), combined with a dedicated energy management strategy (EMS) is discussed. In particular, the mixed H$_2$-HEV architecture is investigated and compared to a series H$_2$-HEV, a parallel H$_2$-HEV, and a base H$_2$-vehicle, which is only equipped with an H$_2$ICE. For hybrid vehicles, a low H$_2$ consumption and low NO$_x^{\mathrm{eo}}$ emissions are conflicting objectives, the trade-off of which depends on the EMS and can be represented as a Pareto front. The dynamic programming algorithm is used to calculate the Pareto-optimal EMS calibrations, to display the H$_2$-NO$_x^{\mathrm{eo}}$ trade-off for various different driving missions. Overall, through the utilization of a dedicated energy management calibration, the mixed H$_2$-HEV demonstrates the capability to consistently achieve extremely low engine-out NO$_x^{\mathrm{eo}}$ emissions across all investigated driving scenarios. For a broad range of driving missions, the mixed H$_2$-HEV is able to decrease the engine-out NO$_\mathrm{x}^\mathrm{eo}$ emissions by more than 90\%, while, at the same time, the H$_2$ consumption is decreased by over 16\%, compared to a comparable non-hybridized H$_2$-vehicle. These significant emission reductions are possible without having to modify the exhaust-gas aftertreatment system, or the optimization of any of the individual drivetrain components, but solely by setting the EMS calibration accordingly.
\end{abstract}


\end{frontmatter}

\section{Introduction} \label{sec:Introduction}
A major challenge of today's society is to decrease the world-wide CO$_2$ emissions. In order to support this transition to a cleaner and more sustainable energy future, renewable energies play a critical role \cite{parra2019review}. However,  with an increasing share of renewable energies, fluctuations in power generation can lead to grid instabilities as energy production does not always align with the current demand. One possibility to cope with the intermittent nature of renewable energies is the use of an energy storage system, which is a carrier that can convert the stored energy back to electric power at any time. Hydrogen energy storage systems are perceived as a relatively inexpensive way of storing, transporting, and trading renewable energies \cite{abe2019hydrogen}, \cite{colbertaldo2019impact}.\\
\\
One major part of the world-wide CO$_2$ emissions stems from the transportation sector. As in 2022, it was accountable for almost one fourth of the global CO$_2$ emissions \cite{IEA_article}, a major part of the global decarbonization politics includes ever more stringent emissions legislations in this sector. In December 2021, the united states environmental protection agency released the final rule \cite{EPA}, setting CO$_2$ emissions standards to \mbox{82.5 g/km} for passenger cars until 2026. The European Union has legally regulated the CO$_2$ emissions of new vehicles to \mbox{95 g/km} until 2025, to \mbox{60 g/km} until 2030, and to \mbox{0 g/km} until 2035 \cite{regulation2019regulation}. This signifies the end of the Diesel and gasoline engines' roles as the dominant technological choices for the light duty market in the EU. However, currently, the traditional combustion engine is still the predominant propulsion technology. In 2019 the global electric-light-vehicles sales accounted for only 2.5\% \cite{gersdorf2020mckinsey}. Therefore, the need to find a replacement for the traditional combustion engine incentivices investments into alternative propulsion technologies that can comply with the net-zero CO$_2$ goal.\\
\\
Hydrogen, which results from renewable energies, has long been known to be used as a CO$_2$-free energy carrier for vehicles. It can be utilized in two ways: Firstly, hydrogen can be used in fuel cells to generate an electric current, which is used to power an electric motor, or stored in a battery. One particular drawback of fuel cells is that they are still rather costly at the moment and require a very high degree of hydrogen purity, since even a low amount of impurity may decrease the fuel cell's performance and irreversibly shorten its lifespan \cite{du2021review}. Secondly, hydrogen can be directly burned in a hydrogen combustion engine (H$_2$ICE). The following sections present a literature review that discusses the possibility of H$_2$ICEs to meet future emission regulations, and explores the options to fill the gap, which will be left by conventional combustion engines.

\subsection{Literature Review on H$_2$ICE}
With the increasing focus on hydrogen as a clean and sustainable energy source, coupled with the necessity to explore alternatives to conventional combustion engines, the H$_2$ICE has emerged as a promising technology for the future. Authors like Shinde et. al. \cite{shinde2022recent}, or Falfari et. al. \cite{falfari2023hydrogen} even propose the H$_2$ICE as the most immediate solution for the near future, as one of its main attractivenesses is that it takes advantage of the current advanced state of ICE technologies. This includes for instance the reliability and durability of combustion engines. Onorati et. al. \cite{onorati2022role} argue that the H$_2$ICE can serve, not only as a temporary fast-transition solution, but also as a long-term solution: They emphasize that already existing supply chains and manufactures, plus the already present recycling infrastructure, make the H$_2$ICE a widespread solution to accelerate the large-scale introduction of H$_2$ICEs into the transportation market. The transferability of knowledge from conventional combustion engines was demonstrated in multiple publications: The authors of \cite{stkepien2021comprehensive} review the applicability of H$_2$ as a fuel for traditional internal combustion engines and conclude that it can be used in both spark ignition as well as compression ignition engines without any major modifications to the system. Additionally, in  \cite{rouleau2021experimental} the adaption of a single cylinder gasoline engine is described for hydrogen combustion and a peak efficiency of up to 47\% could be achieved, highlighting the technology's excellent combustion efficiency.\\
\\
For H$_2$ICEs to be considered a viable future propulsion system, they must also demonstrate compliance with forthcoming greenhouse gas emissions regulations. The new \mbox{EURO 7} norm for light-duty vehicles is to be expected to become effective in 2025 and includes a fuel-neutral limit for tailpipe NO$_\mathrm{x}$ emissions (from now on abbreviated with NO$_\mathrm{x}^\mathrm{tp}$) of \mbox{60 mg/km} \cite{european2021annexes}. Hydrogen internal combustion engines have long been known for the ability to burn hydrogen with extremely low engine-out NO$_\mathrm{x}$ emissions (from now on abbreviated with NO$_x^\mathrm{eo}$) \cite{white2006hydrogen}. Different strategies are known to achieve such low NO$_x^\mathrm{eo}$ emissions during the combustion process of hydrogen. One way is to decrease the combustion temperature, which can be achieved for instance by exhaust gas recirculation (EGR) where already burnt gases are fed back and used in the next combustion. Heffel presents in \cite{heffel2003nox_1}, \cite{heffel2003nox_2} the use of EGR to reduce the NO$_x^\mathrm{eo}$ formation of an H$_2$ICE. Although he could show that the NO$_x^\mathrm{eo}$ formation can be pushed below 10 ppm under stoichiometric conditions, most new publications discuss to exploit the wide flammability range of H$_2$ to reduce NO$_x^\mathrm{eo}$ emissions. This characteristic allows to burn hydrogen using a large surplus of air to decrease the combustion temperature, which is known as ultra-lean combustion. Multiple groups have reported extremely low NO$_x^\mathrm{eo}$ emissions using air-to-fuel ratios up to $\lambda\approx 3$. Bao et. al. \cite{bao2022experimental} show that the utilization of a turbocharger and improved injection pressure increase the H$_2$ICE's power and efficiency simultaneously. This way they show that extremely low NO$_x^\mathrm{eo}$ emissions can be reached with a maximum brake thermal efficiency of 40.4\%; Sementa et. al. \cite{sementa2022exploring} investigated different H$_2$ dilutions in experiments and achieved stable combustion up to $\lambda = 3.4$, presenting no CO emissions, very low HC emissions, originating from the lubricant oil, and extremely low NO$_x^\mathrm{eo}$ emissions.\\
\\
Whereas ultra-lean combustion can result in excellent emission characteristics, it also has the major drawback of a limited power density compared to stoichiometric or rich combustion conditions. To achieve a high engine power output, increasingly more H$_2$ has to be injected, which ultimately leads to a lower air-to-fuel ratio. Two issues are to be expected in this situation: Firstly, since ultra-lean combustion is not possible anymore, increasing NO$_x^\mathrm{eo}$ emissions will occur, partly nullifying the H$_2$ICE's advantage over conventional combustion engines. Secondly, at lower air-to-fuel ratios, H$_2$ICEs are known to exhibit combustion irregularities such as spontaneous ignition, or knocking \cite{falfari2023hydrogen}. This can be challenging for day-to-day applications in a car, as high engine-load operations are commonly encountered, e.g., acceleration in urban areas, or high-speed driving on the highway. However, if such a vehicle would additionally be equipped with a secondary torque source, e.g., an electric motor, theoretically, both issues could be tackled by assisting the H$_2$ICE during high engine-loads. In the following, the so-called energy management system (EMS) for hybrid electric vehicles (HEVs) is explained and (if handled properly) its positive effect on reducing the emissions is outlined. Additionally, a summary of the existing literature on HEVs that use an H$_2$ICE is presented.
\subsection{Hybrid Electric Vehicles}
The EMS is a necessary control algorithm for every HEV and stems from the fact that any HEV has more than one source to produce power. The EMS is concerned about the provision of the so-called driver's power request, which is (given the vehicle speed and road inclination) a consequence of the driver's desired vehicle acceleration. Assuming for instance that the driver demands \mbox{20 kW}, it remains ambiguous whether this power should originate from the combustive part, the electric part, or a combination of both. This decision of the power allocation is referred to as the power split. Sciaretta et. al have shown in \cite{sciarretta2014control} the importance of an elaborate EMS algorithm. The authors compared nine different EMS algorithms for the same vehicle and reported that the best algorithm decreased the overall CO$_2$ emissions by 28\%.\\
Early approaches for the EMS were of heuristic nature, i.e., they consist of rules which are based on expert's experience to decide the power split. State-of-the-art algorithms often are based on solving an optimization problem, which results of a mathematical description of the power allocation problem. Such optimization problems are called optimal control problems (OCPs) and include the formulation of mathematical constraint functions, which describe the HEV powertrain sufficiently well, and one (or possibly more) quality criterion for each realizable distribution of the power request. Typical quality criteria would be to achieve minimal CO$_2$ emissions (e.g. \cite{ritzmann2019fuel}), or the prolongation of battery life (e.g. \cite{serrao2011optimal}). The goal of optimization-based EMS algorithms is to find the optimal power split by solving such an OCP using dedicated mathematical solvers.\\
Overall, EMS control algorithms can be divided into two distinctively different categories, i.e., online controllers, and offline algorithms. Online controllers are designed such that they can be implemented on the vehicle's embedded control unit. They are causal, meaning that all signals necessary for calculating the controller outputs are accessible at the instance of the computation process.  Here, the main challenges are to achieve a good trade-off between computing a power split, which leads to acceptable CO$_2$ emissions, and complying with the restrictions of the embedded hardware on the vehicle. Offline EMS algorithms, however, are simulative investigations of the potential of the HEV's fuel saving capabilities. They are often described as non-causal, meaning that they have access to all constraints and disturbances of the OCP at each point in time. Here, the goal is to find the theoretically best possible solution to the OCP, which describes finding the HEVs optimal power split at every point in time for a predefined driving mission. Although such investigations cannot capture all effects that occur during real-world driving, they are a valuable tool to investigate the impact of different powertrain topologies on the achievable performance in simulation before building the physical system.\\
\\
One well-known extension of the energy management problem includes the NO$_\mathrm{x}$ emissions. As a result, a typical multi-objective optimization problem is obtained, which means that the OCP includes two competing objectives, namely: the minimization of CO$_2$ emissions, and the minimization of NO$_\mathrm{x}$ emissions. Since these objectives are in conflict, such OCPs do not have one single optimal solution anymore, but a multitude of solutions that lie on the so-called Pareto front. Each solution on this Pareto front is a different optimal realization of the fuel-NO$_\mathrm{x}$ trade-off, i.e., reducing CO$_2$ always comes at the expense of increasing NO$_\mathrm{x}$, and vice-versa. Using offline algorithms, Ritzmann et. al. have investigated the fuel-NO$_\mathrm{x}$ trade-off potential of a parallel HEV that is equipped with a Diesel engine, and compared it to a non-electrified vehicle \cite{ritzmann2021optimal}. They have shown that for a comparable fuel consumption, the NO$_\mathrm{x}^\mathrm{tp}$ can roughly be halved by using the hybrid propulsion technology. Moreover, the fuel-NO$_\mathrm{x}$ trade-off that results from different choices of the EMS controller design, spans a wide range.\\
\\
Although the literature is rich in examples for experiments addressing HEVs that are equipped with either a Diesel engine or a gasoline engine, there are only few reports on vehicles that use hydrogen combustion as part of their HEV architecture (from herein on referred to as H$_2$-HEV). The only two test vehicles presented in the literature are as follows: In \cite{he2006development}, a standard sport utility vehicle has been converted into a hydrogen-powered H$_2$-HEV. In \cite{jaura2004ford}, a hydrogen engine propelled hybrid electric concept vehicle, called Ford's H$_2$RV, is described. In both studies, the authors state excellent NO$_\mathrm{x}^\mathrm{tp}$ emissions while achieving low H$_2$ consumption. However, the achieved results would not satisfy the future legislation limitations that are expected with the introduction of, e.g., \mbox{EURO 7}. Two reasons can be identified for the relatively high NO$_\mathrm{x}^\mathrm{tp}$ emissions: Firstly, both studies are based on old examples of vehicles (2006 and 2004). Secondly, both vehicles were equipped with EMS algorithms that are inferior to todays state-of-the-art methods, which can have a substantial influence on the fuel-NO$_\mathrm{x}$ trade-off (from now on referred to as the H$_2$-NO$_\mathrm{x}$ trade-off).\\
\\
Only two publications are found in the literature regarding the simulative performance of an H$_2$-HEV that is equipped with a modern turbocharged H$_2$ICE. The first publication, written by Beccari et. al. \cite{beccari2022use}, compares the CO$_2$ emissions of a standard HEV with a Diesel engine to an H$_2$-HEV. Although not elaborated in detail, the authors simply describe to operate the combustion engine in its optimal operating point at each point in time. However, this does neither explain how the electrical components are operated, nor are any effects on the battery energy content shown and discussed. Additionally, the NO$_\mathrm{x}$ emissions are not taken into consideration and the investigations are limited to one driving cycle.\\
The second publication, written by Kyjovsk\'y et. al. \cite{kyjovsky2023drive}, compares the CO$_2$ and NO$_\mathrm{x}$ emissions of parallel H$_2$-HEVs from different weight classes on the WLTC in simulation. They report good fuel consumption, while the lighter vehicles show NO$_\mathrm{x}$ emissions below the limits imposed by EURO 6, without the use of an aftertreatment system. As the main focus of this publication lies in the investigation of turbocharging and engine downsizing, the EMS plays a subordinated role. Therefore, instead of computing the entire realizable Pareto front, a standard EMS algorithm is implemented, which results in a sub-optimal vehicle performance. As a result, the approach presented has two major shortcomings: Firstly, no measure is given of how much CO$_2$ (or NO$_\mathrm{x}$, respectively) can be saved using a theoretically optimal EMS controller. Secondly, without the full Pareto front, it is impossible to know how much either of these two quantities could have been reduced at the expense of the other.
\subsection{Contribution}
In this publication, a simulative comparison between three different H$_2$-HEV architectures and a base H$_2$-vehicle, which is only equipped with an H$_2$ICE, is conducted. The different hybrid propulsion architectures feature a series H$_2$-HEV, a parallel H$_2$-HEV, and a mixed H$_2$-HEV. The contribution of this publication is twofold.
\begin{enumerate}
	\item To the authors' best knowledge, for the first time, this publication presents the full performance potential that is achievable by an optimal EMS calibration on different H$_2$-HEVs. This includes the entire achievable trade-off between H$_2$ consumption and engine-out NO$_\mathrm{x}^\mathrm{eo}$ emissions. These H$_2$-NO$_\mathrm{x}^\mathrm{eo}$ Pareto fronts are obtained with the dynamic programming algorithm (DP), which guarantees global optimality for a predefined driving mission. A comparison on the WLTC reveals that all hybrid propulsion architectures outperform the non-hybridized H$_2$ICE, whereas the mixed H$_2$-HEV is superior over the series and the parallel configuration.
	\item To solidify the findings, the achievable trade-off between H$_2$ consumption and engine-out NO$_\mathrm{x}^\mathrm{eo}$ emissions of the mixed H$_2$-HEV is presented for a wide variety of driving missions. An in-depth analysis of different realizations of H$_2$-NO$_\mathrm{x}^\mathrm{eo}$-optimal solutions provides insight into how the mixed H$_2$-HEV architecture can be exploited to operate the H$_2$ICE solely under ultra-lean conditions, resulting in extremely low NO$_\mathrm{x}^\mathrm{eo}$ emissions. Consequently, two worst-case driving scenarios are analyzed to highlight the technical limitations of the mixed H$_2$-HEV. In conclusion, to the authors' best knowledge, this study provides the first complete overview of the potential of H$_2$-HEVs to satisfy future emission legislations, solely via the optimal calibration of the EMS algorithm.
\end{enumerate}
The remainder of this paper is organized as follows: In \mbox{Section \ref{sec:ProblemDescription}}, the different H$_2$-HEV models are outlined and the energy management problem including NO$_\mathrm{x}^\mathrm{eo}$ emissions is formulated as an OCP. In \mbox{Section \ref{sec:Results}}, a comparison between the performance of the different H$_2$-HEVs and the base H$_2$-vehicle is presented on the WLTC. Furthermore, with regards to the H$_2$-NO$_\mathrm{x}^\mathrm{eo}$ trade-off, the benefits of hybridization over only using an H$_2$ICE are discussed and the superiority of the mixed H$_2$-HEV is explained. In \mbox{Section \ref{sec:different_driving_missions}}, the achievable H$_2$-NO$_\mathrm{x}^\mathrm{eo}$ trade-off of the mixed H$_2$-HEV is presented for four additional driving missions, including two worst-case driving scenarios, highlighting the technology's limitations. In \mbox{Section \ref{sec:Conclusion}}, an outlook on future research is presented.

\section{Problem Description} \label{sec:ProblemDescription}
In this section, the four investigated drivetrain configurations are introduced. Firstly, the base H$_2$-vehicle is explained, then the three hybridized H$_2$-HEVs are presented. Subsequently the energy management problem including NO$_\mathrm{x}^\mathrm{eo}$ emissions is outlined.
\subsection{Drivetrain Configurations}
\begin{figure}[h!]
\centering
   \def\svgwidth{1\columnwidth}
{\normalsize
	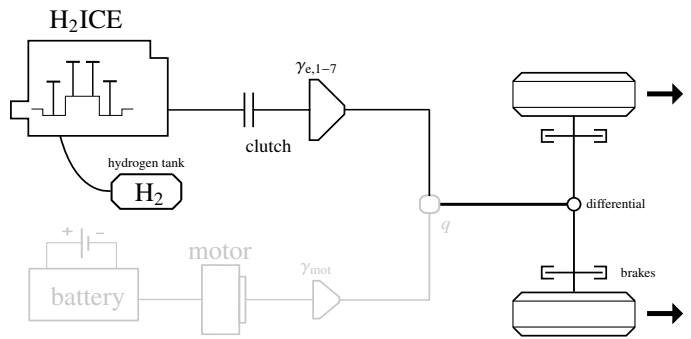}
    \caption{Schematic of the base H$_2$-vehicle and the parallel H$_2$-HEV. The components that are grayed out are only part of the parallel H$_2$-HEV.}
    \label{fig:parallelDrivetrain}
\end{figure}

\begin{figure}[h!]
\centering
   \def\svgwidth{1\columnwidth}
{\normalsize
	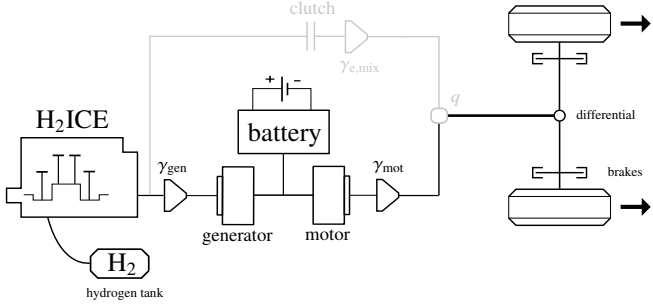}
    \caption{Schematic of the base series H$_2$-HEV and the mixed H$_2$-HEV. The components grayed out are only part of the mixed H$_2$-HEV.}
    \label{fig:seriesDrivetrain}
\end{figure}

For the sake of comparison, the investigated H$_2$-HEVs are built upon the base H$_2$-vehicle, but with additional electrical components. This ensures uniformity across the analyzed vehicles with respect to the H$_2$ICE, aerodynamic drag and rolling resistance, and the base H$_2$-vehicle's mass $m_\mathrm{base}$. The H$_2$-HEVs have additional weight, according to their individual electrical equipment. Table \ref{tab:vehicleComponents} summarizes the individual vehicle parameters. Note that the vehicle components are additive, meaning that, if, e.g., the series H$_2$-HEV is considered, then it also includes the components of the base H$_2$-vehicle and the parallel H$_2$-HEV.\\
\\
For all investigated propulsion architectures, the engine ON/OFF state is described by the binary variable $e_0$ and the clutch state (if a clutch is included) is described by the binary variable $c_0$. While a value of $e_0=1$ indicates that the engine is running, $e_0=0$ indicates that it is turned off. Regarding the clutch, $c_0=1$ indicates that the clutch is engaged, while $c_0=0$ indicates that the clutch is open.\\
\\
\textit{Base vehicle:}\\
Figure \ref{fig:parallelDrivetrain} shows a schematic of the powertrain architecture of the base H$_2$-vehicle and electric parts in gray. The electric parts are only used in the parallel H$_2$-HEV drivetrain and are introduced in a section further down.%
The H$_2$ICE is connected to the front axle through a seven speed gearbox, the clutch, and a differential. The seven speed gearbox's transmission ratios are denoted by $\gamma_{\mathrm{e,1-7}}$. As for this vehicle the only power source is the H$_2$ICE, any positive torque request of the driver has to be supplied by the H$_2$ICE. If the torque request is negative, or zero, then the clutch is opened ($c_0=0$), the engine is shut-off ($e_0=0$), and the friction brakes are used. As a result, engine idling is prevented. Engine braking is not considered in this work.\\
\\
\textit{Parallel H$_2$-HEV:}\\
The parallel H$_2$-HEV propulsion architecture results from adding the electrical parts that are grayed out in Figure \ref{fig:parallelDrivetrain}. A torque split device, which is denoted by $q$, is used before the gearbox, resulting in a P3 hybrid. The additional electric motor uses the fixed-gear transmission $\gamma_\mathrm{mot}$ and can draw power from or feed power to the battery. In addition to the base weight of the vehicle, the electric devices account for \mbox{$m_\mathrm{e,par}=160$ kg}, which includes the battery, an electric motor, and power electronics.\\
\\
\textit{Series H$_2$-HEV:}\\
Figure \ref{fig:seriesDrivetrain} shows a schematic of the powertrain architecture of the series H$_2$-HEV and grayed out parts. The grayed out clutch, fixed gearbox, and torque split device are used in the mixed H$_2$-HEV drivetrain and are introduced in a section further down. The H$_2$ICE is not physically connected to the wheels, but is only used to produce electric energy via the generator. This electric energy is either directly used by the motor, fed to the battery, or employed in both ways simultaneously. The motor and the generator are the same make and model, whereas the motor has the same fixed transmission ratio as in the parallel H$_2$-HEV, and the generator has a different fixed transmission ratio $\gamma_\mathrm{gen}$. In addition to the base weight of the vehicle, the electric devices account for $m_\mathrm{e,ser}=200$ kg, which includes the battery, the electric motor, the generator, and power electronics.
\\
\textit{Mixed H$_2$-HEV:}\\
The mixed H$_2$-HEV propulsion architecture results from adding the parts in Figure \ref{fig:seriesDrivetrain} that are grayed out. The clutch can be used to connect the H$_2$ICE to the wheels via a fixed transmission $\gamma_\mathrm{e,mix}$ and the torque split device $q$. This drivetrain architecture is called mixed H$_2$-HEV, since it combines a series H$_2$-HEV and a parallel H$_2$-HEV architecture. In addition to the base weight of the vehicle, the electric devices account for $m_\mathrm{e,mix}=200$ kg, which includes the battery, the electric motor, the generator, and power electronics.\\
\\
\textit{Driving modes:}\\
Hybrid electric vehicles encompass various driving modes that arise from the flexibility to either use multiple power sources simultaneously, or to rely on only one power source and to decouple, or switch off, the other one(s). The different driving modes are denoted by $M\in \{1,2,3\}$ in this work, and are listed in Table \ref{tab:drivingModes}. In series mode, the H$_2$ICE is running, but it is decoupled from the wheels, i.e., any requested power is fully delivered by the motor. In parallel mode, the H$_2$ICE is running and physically coupled to the wheels. In this mode, both propulsion systems (H$_2$ICE and motor) can provide power to the wheels. It also enables load-point shifting, where the combustion engine produces more power than is required at the wheels and the access power is stored in the battery. In EV mode, the combustion engine is off and decoupled from the wheels. The entire torque request is met by the motor.
\begin{table}[h!]
\vspace{-0.2cm}
\centering
\caption{Vehicle parameters of the four propulsion architectures.}
\label{tab:vehicleComponents}
\begin{tabular}{l| l| l}\hhline{-|-|-}
\multirow{3}{*}{base H$_2$-vehicle} & weight & 1300 kg\\
 & \multirow{2}{*}{H$_2$ICE} & 4-cyl\\
 &  & 163 kW\\\hhline{-|-|-}
\multirow{5}{*}{parallel H$_2$-HEV} & \multirow{2}{*}{motor} & 173 kW\\
 &  & 40 kg \\\cline{2-3}
 & \multirow{3}{*}{battery} & 230 V \\
 &  & 11 kWh \\
 &  & 120 kg \\\hhline{-|-|-}
\multirow{2}{*}{series- \& mixed H$_2$-HEV} & \multirow{2}{*}{generator} & 173 kW\\ 
 &  & 40 kg \\\hhline{-|-|-}
\end{tabular}
\end{table}
\begin{table}[h!]
\vspace{-0.2cm}
\centering
\caption{Driving modes of the different propulsion architectures.}
\label{tab:drivingModes}
\begin{tabular}{l| l| l}\hhline{-|-|-}
Series mode & $c_0=0$, $e_0=1$ & $M=1$\\
Parallel mode & $c_0=1$, $e_0=1$ & $M=2$\\
EV mode & $c_0=0$, $e_0=0$ & $M=3$\\\hline
\end{tabular}
\end{table}
\subsection{Drivetrain Model} 
\begin{figure}[h!]
    \centering
    \includegraphics[scale=0.4]{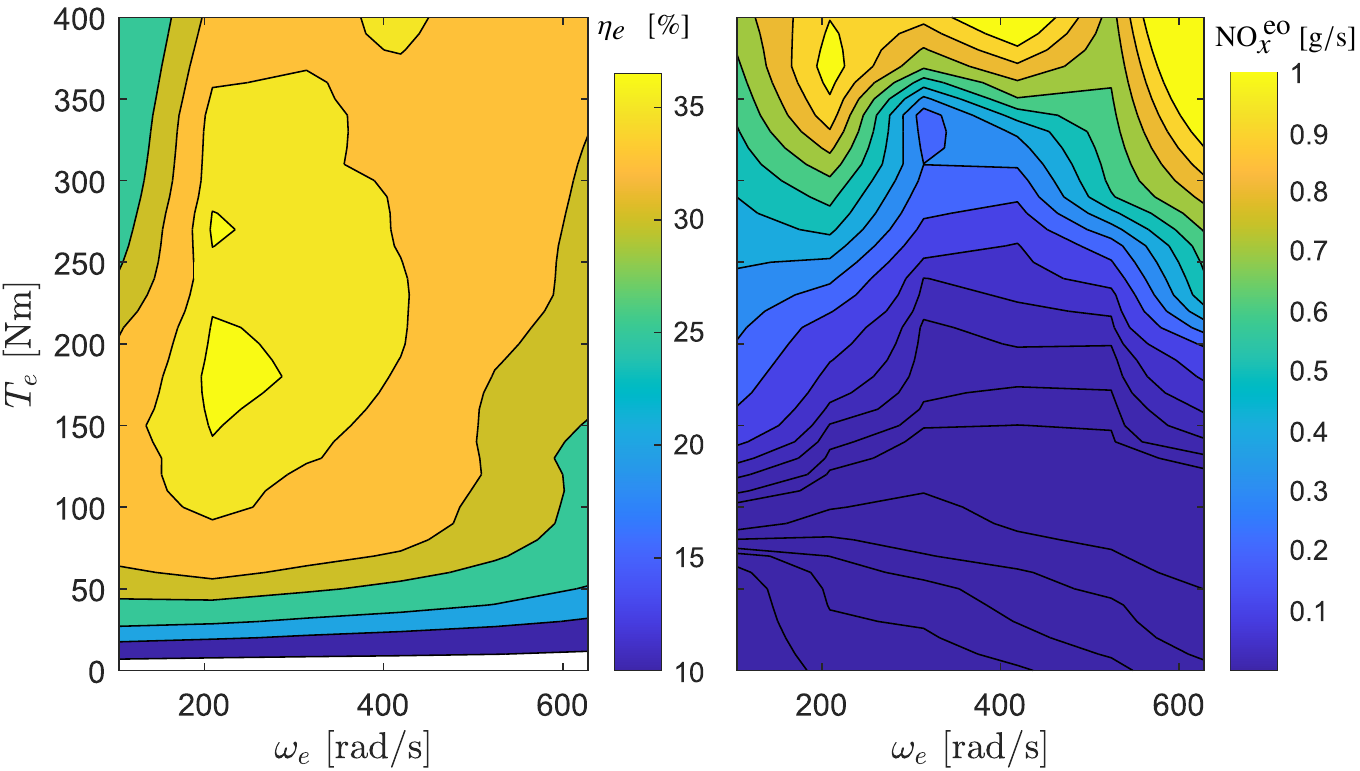}
    \caption{Left: H$_2$ICE efficiency map. Right: NO$_\mathrm{x}^\mathrm{eo}$ map.}
    \label{fig:fuelNOx}
\end{figure}

\begin{figure}[h!]
    \centering
    \includegraphics[scale=0.7]{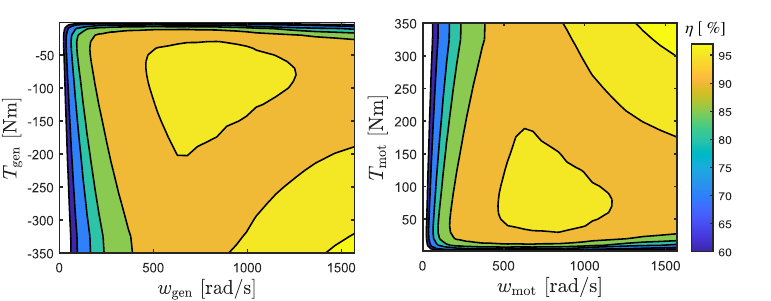}
    \caption{Left: Generator efficiency map. Right: Motor efficiency map. Motor and generator are of the same make and model.}
    \label{fig:lambda}
\end{figure}

The modeling approach used in this work is based on \cite{guzzella2013vehicle} and is outlined in the following.\\
\\
\textit{Propulsion systems:}\\
The engine's fuel consumption and the NO$_\mathrm{x}^\mathrm{eo}$ emissions are obtained using steady-state mappings that were evaluated on a test bench and depend on the engine speed $\omega_{e}$ and the engine torque $T_{e}$:
\begin{align} 
\dot{m}_f &= f(\omega_{e}, T_e) \label{eq:m_fuel_dot}\\
\dot{m}_{\mathrm{NO}_\mathrm{x}} &= f(\omega_{e}, T_e). \label{eq:m_NOx_dot}
\end{align}
The obtained maps are depicted in Figure \ref{fig:fuelNOx}. Extremely low NO$_\mathrm{x}^\mathrm{eo}$ can be reached under low-load conditions. This is possible as the turbocharged H$_2$ICE is  calibrated to operate under ultra-lean combustion conditions in this region, leading to NO$_\mathrm{x}^\mathrm{eo}$ emissions below 2 mg/s.
\\
The electric motor and generator losses are stored in maps that are derived from high-fidelity component models:
\begin{align}
	P_{\mathrm{l}_\mathrm{mot}} = f(T_{\mathrm{mot}}, \omega_\mathrm{mot}),\\
	P_{\mathrm{l}_\mathrm{gen}} = f(T_{\mathrm{gen}}, \omega_\mathrm{gen}).
\end{align}
They depend on the torque and the rotational speed of the corresponding component. Their efficiency maps are depicted in Figure \ref{fig:lambda}. 
\\
\textit{Torque request:}\\
The requested force at the wheels is a function of the aerodynamic drag force and rolling resistance $F_d$, the gravitational force $F_g$, and the inertial force $F_i$ to accelerate the vehicle:
\begin{equation} \label{eq:vehiclePower}
	F_{\mathrm{req}} = F_d(v) + F_g(\Gamma) + F_i(a),
\end{equation}
where $\Gamma$ denotes the road gradient, $v$ the velocity, and $a$ the vehicle acceleration. The resistive force $F_d$ is approximated using a quadratic dependency on $v$. The inertial force $F_i$ is calculated by accelerating the total mass of the vehicle $m_\mathrm{tot}$:
\begin{equation} \label{eq:mtot}
	\begin{aligned}
		F_i &= m_\mathrm{tot} \cdot a,\\
		m_\mathrm{tot} &= 
			\begin{cases}
				m_\mathrm{base} + m_\mathrm{e,par} & \text{,if  parallel H$_2$-HEV},\\
				m_\mathrm{base} + m_\mathrm{e,ser} & \text{,if  series H$_2$-HEV},\\
				m_\mathrm{base} + m_\mathrm{e,mix} & \text{,if  mixed H$_2$-HEV},
			\end{cases}
	\end{aligned}
\end{equation}
where $m_{\mathrm{e,}j}, j \in \{\mathrm{par}, \mathrm{ser}, \mathrm{mix}\}$ accounts for the sum of all additional electric devices that make up the hybridized versions of the base H$_2$-vehicle. The resulting torque request is calculated using the wheel's radius $r_w$:
\begin{equation} \label{eq:force_to_power}
	T_{\mathrm{req}} = F_{\mathrm{req}} \cdot r_w.
\end{equation}
The torque request has to be provided by the vehicle's combined drivetrain. For the case of the base H$_2$-vehicle, the entire torque is delivered by the H$_2$ICE, which means that the H$_2$ consumption and NO$_\mathrm{x}^\mathrm{eo}$ emissions can directly be calculated using (\ref{eq:m_fuel_dot}) and (\ref{eq:m_NOx_dot}). For the case of the H$_2$-HEVs, the torque request can be divided between multiple power sources at each point in time. This power distribution is explained in the following for the mixed H$_2$-HEV. However, the torque splits for the parallel H$_2$-HEV and the series H$_2$-HEV can directly be obtained from the same description.\\
\\
\textit{Rotational Speeds:}\\
Before the torque split is introduced, the description of the rotational speeds of the mixed H$_2$-HEV components are explained. From the vehicle speed, the wheel speed calculates as:
\begin{equation} \label{eq:omega_w}
\omega_w = \frac{v}{r_w}.
\end{equation}
As the motor is always physically coupled to the wheels, its speed is directly defined by:
\begin{equation} \label{eq:omega_mot}
\omega_\mathrm{mot} = \omega_w \cdot \gamma_\mathrm{fd} \cdot \gamma_\mathrm{mot},
\end{equation}
whereas $\gamma_\mathrm{fd}$ and $\gamma_\mathrm{mot}$ are the fixed gear ratios of the final drive, and the motor, respectively. The engine is not always physically coupled to the wheels, as this connection depends on the driving mode. If the vehicle is in parallel mode, $\omega_\mathrm{e}$ is defined by the wheel speed and the respective engine transmission ratio $\gamma_{e,i}$. The engine transmission is different for the individual vehicles, which is here denoted by the subscript $i \in \{\mathrm{mix}, \mathrm{1-7}\}$. The mixed H$_2$-HEV's gear ratio $\gamma_\mathrm{e,mix}$ is fixed, whereas for the base H$_2$-vehicle and for the parallel H$_2$-HEV, $\gamma_\mathrm{e,1-7}$ is defined by the seven-speed gearbox, which again is encoded via the vehicle speed. However, if the vehicle is in series mode, or in EV mode, then the clutch is open and $\omega_\mathrm{e}$ is defined by the generator speed $\omega_\mathrm{gen}$:
\begin{equation} \label{eq:omega_e}
	\omega_\mathrm{e} = 
		\begin{cases}
			\omega_w \cdot \gamma_\mathrm{fd} \cdot \gamma_{\mathrm{e,i}}& \text{,if $M = 2$},\\
			\frac{\omega_\mathrm{gen}}{\gamma_\mathrm{gen}} & \text{,if $M \in \{1,3\}$},
		\end{cases}
\end{equation}
whereas $\gamma_\mathrm{gen}$ is the fixed gear ratio of the generator. \\
\\
Similar to the engine speed, the generator speed depends on the driving mode. In parallel mode, $\omega_\mathrm{gen}$ is fixed by the vehicle speed. In EV mode, the generator is switched off by design and its rotational speed is zero. In series mode, the generator speed can be chosen freely and does not depend on the vehicle speed.
\begin{equation} \label{eq:omega_gen}
	\omega_\mathrm{gen} = 
		\begin{cases}
			\in [\omega_\mathrm{gen}^\mathrm{min}, \, \omega_\mathrm{gen}^\mathrm{max}] & \text{,if $M = 1$},\\
			\omega_w \cdot \gamma_\mathrm{fd} \cdot \gamma_\mathrm{gen} & \text{,if $M = 2$},\\
			0 & \text{,if $M = 3$}.
		\end{cases}
\end{equation}
Here, $\omega_\mathrm{gen}^\mathrm{min}$ and $\omega_\mathrm{gen}^\mathrm{max}$ represent physical limits implied by the engine idle speed and the engine max speed.\\
\\
\textit{Torque Split:}\\
After all rotational speeds are defined, the torque split for the mixed H$_2$-HEV can be formulated. Including the efficiency of the final drive $\eta_\mathrm{fd}$, the torque request is calculated before the torque split device as:
\begin{equation} \label{eq:Treq_diff}
	T_{\mathrm{req}_\mathrm{fd}} = 
		\begin{cases}
			\nicefrac{T_\mathrm{req}}{\gamma_\mathrm{fd}}  \div \eta_\mathrm{fd} & \text{,if $T_\mathrm{req} \geq 0$},\\
			\nicefrac{T_\mathrm{req}}{\gamma_\mathrm{fd}} \,\cdot \, \eta_\mathrm{fd} & \text{,if $T_\mathrm{req} < 0$}.
		\end{cases}
\end{equation}
Considering the motor gearbox efficiency $\eta_{\mathrm{GB}_\mathrm{mot}}$, the torque split device $q$ is used to distribute $T_{\mathrm{req}_\mathrm{fd}}$ between the electric motor torque:
\begin{equation} \label{eq:Tmot}
	\begin{aligned}
		T_{\mathrm{mot}_\mathrm{req}} &= T_{\mathrm{req}_\mathrm{fd}} \cdot q,\\
		T_{\mathrm{mot}} &= 
			\begin{cases}
				\frac{T_{\mathrm{mot}_\mathrm{req}}}{\gamma_{\mathrm{mot}}} \div \eta_{\mathrm{GB}_\mathrm{mot}} & \text{,if $T_{\mathrm{req}_\mathrm{fd}} \geq 0$},\\
				\frac{T_{\mathrm{mot}_\mathrm{req}}}{\gamma_{\mathrm{mot}}} \, \cdot \, \eta_{\mathrm{GB}_\mathrm{mot}} & \text{,if $T_{\mathrm{req}_\mathrm{fd}} < 0$},
			\end{cases}
	\end{aligned}
\end{equation}
and the torque that is requested from the combined engine-generator-unit:
\begin{equation} \label{eq:Te_req}
	T_{e_\mathrm{req}} = T_{\mathrm{req}_\mathrm{fd}} \cdot (1-q).
\end{equation}
In series mode and in EV mode, the full power is delivered by the motor. But in parallel mode, the torque split can be chosen arbitrarily as long as the torque distribution can be delivered by the corresponding components. Therefore, the possible choice of torque splits is defined as:
\begin{equation} \label{eq:torqueSplit}
	q = 
		\begin{cases}
			1 & \text{,if $M \in \{1,\, 3\}$}\\
			]-\infty, \infty[ & \text{,if $M = 2$}.
		\end{cases}
\end{equation}
As the generator and the engine are physically coupled, the final torque, which the engine has to deliver, is defined by the generator gearbox efficiency $\eta_{\mathrm{GB}\mathrm{gen}}$ and the generator torque:
\begin{equation} \label{eq:T_e}
	T_{e} = T_{e_\mathrm{req}} - \frac{T_{\mathrm{gen}} \cdot \gamma_{\mathrm{gen}}}{\eta_{\mathrm{GB}\mathrm{gen}}}.
\end{equation}
Similar to the torque split, the choice of obtainable generator torques depends on the driving mode:
\begin{equation} \label{eq:generatorTorque}
	T_\mathrm{gen} = 
		\begin{cases}
			[T^\mathrm{min}, \, T^\mathrm{max}] & \text{$M \in \{1,\, 2\}$}\\
			0 & \text{$M = 3$}.
		\end{cases}
\end{equation}
The torque bounds $T^\mathrm{min}$ and $T^\mathrm{max}$ represent the torque limits of the combined engine-generator unit. 
Finally, to calculate the electric power that is drawn from the battery, the source power that is provided by the motor and generator calculates as:
\begin{equation} \label{eq:Pmot_source}
	P_{\mathrm{mot}_\mathrm{s}} = T_{\mathrm{mot}} \cdot \omega_\mathrm{mot} + P_{\mathrm{l}_\mathrm{mot}},
\end{equation}
and
\begin{equation} \label{eq:Pgen_source}
	P_{\mathrm{gen}_\mathrm{s}} = T_{\mathrm{gen}} \cdot \omega_\mathrm{gen} + P_{\mathrm{l}_\mathrm{gen}}.
\end{equation}
The battery is modeled using a Thévenin equivalent model \mbox{\cite{guzzella2013vehicle}}. The current $I_b$ and that results from a requested battery power $P_b$ calculates as
\begin{equation} \label{eq:thevenin1}
	\begin{aligned}
		I_b &= \frac{V_{oc}-\sqrt{V_{oc}^2-4R_i P_b}}{2R_i},\\
		P_b &= P_{\mathrm{mot}_\mathrm{s}}+P_{\mathrm{gen}_\mathrm{s}}+P_{\mathrm{aux}},
	\end{aligned}
\end{equation}
where the inner resistance $R_i$ and the auxiliary losses $P_{\mathrm{aux}}$ are assumed to be constant. The battery source power calculates as
\begin{equation} \label{eq:thevenin2}
	\begin{aligned}
		P_{s_b} &= P_b + P_{l_b},\\
		P_{l_b} &= R_i \cdot I_b^2,
	\end{aligned}
\end{equation}
where the battery internal losses are denoted by $P_{\mathrm{l}_b}$.
The high-voltage battery open circuit voltage $V_{\mathrm{oc}}$ can be approximated to depend linearly on the state of charge (SoC) \cite{widmer2022battery} and is described in this work as:
\begin{equation} \label{eq:Voc}
	V_{\mathrm{oc}} = \alpha_\mathrm{bat} \cdot \mathrm{SoC}+\beta_\mathrm{bat},
\end{equation}
where $\alpha_\mathrm{bat}$, $\beta_\mathrm{bat}$ are fitted parameters.
The battery SoC dynamics are described by:
\begin{equation} \label{eq:soc_dynamics}
	\begin{aligned}
		\dot{\mathrm{SoC}} &=  \frac{-P_{s_b}}{Q_\mathrm{max}\cdot V_\mathrm{oc}},
	\end{aligned}
\end{equation}
where $Q_\mathrm{max}$ is the full battery capacity.\\
\\
Overall, the above introduced torque split description is valid for the mixed H$_2$-HEV architecture. However, the torque splits for the series H$_2$-HEV and the parallel H$_2$-HEV are obtained by choosing the corresponding parts of (\ref{eq:omega_e}), (\ref{eq:omega_gen}), (\ref{eq:torqueSplit}), and (\ref{eq:generatorTorque}).
\subsection{Energy Management Including NO$_\mathrm{x}^\mathrm{eo}$:}
To formulate the optimal control problem of the energy management including NO$_\mathrm{x}^\mathrm{eo}$, the vector of dynamic state variables and its derivative are defined as:
\begin{equation} \label{eq:x_state}
	\begin{aligned}
		\mathbf{x} &= [\mathrm{SoC}, \, m^\mathrm{eo}_{\mathrm{NO}_\mathrm{x}}]^T,\\
		\mathbf{\dot{x}} &= [\dot{\mathrm{SoC}}, \, \dot{m}^\mathrm{eo}_{\mathrm{NO}_\mathrm{x}}]^T = \mathbf{f}(\mathbf{x},\mathbf{u}).
	\end{aligned}
\end{equation}
Here, the accumulated and instantaneous NO$_\mathrm{x}^\mathrm{eo}$ emissions are denoted by $m^\mathrm{eo}_{\mathrm{NO}_\mathrm{x}}$, and $\dot{m}^\mathrm{eo}_{\mathrm{NO}_\mathrm{x}}$, respectively. The variable $\mathbf{u}$ represents the vector of control inputs, which depends on the investigated vehicle. The most general case is for the mixed H$_2$-HEV, as this architecture offers all degrees of freedom to manipulate the torque split. It is defined as:
\begin{equation} \label{eq:u_controlInputs}
	\mathbf{u} = [T_\mathrm{gen}, \, \omega_\mathrm{gen}, \, q, \, M]^T.
\end{equation}
The base H$_2$-vehicle and the remaining H$_2$-HEVs use a subset of $\mathbf{u}$. Table \ref{tab:DOF_architectures} lists an overview of the sets of control inputs for all investigated vehicles.
\begin{table}[h!]
\centering
\caption{Overview of the degrees of freedom of the investigated propulsion architectures.}
\label{tab:DOF_architectures}
\begin{tabular}{l| l| l| l| l}
 & $T_\mathrm{gen}$ & $\omega_\mathrm{gen}$ & $q$ & $M$\\\hhline{-|-|-|-|-}
Base H$_2$-vehicle & $\times$ & $\times$ & $\times$ & $\times$\\\hhline{-|-|-|-|-}
Parallel H$_2$-HEV & $\times$ & $\times$ & $\checkmark$ & $\{2,\,3\}$\\\hhline{-|-|-|-|-}
Series H$_2$-HEV & $\checkmark$ & $\checkmark$ & $\times$ & $\{1,\,3\}$\\\hhline{-|-|-|-|-}
Mixed H$_2$-HEV & $\checkmark$ & $\checkmark$ & $\checkmark$ & $\{1,\,2,\,3\}$\\\hhline{-|-|-|-|-}
\end{tabular}
\end{table}
\\
\\
The goal of solving the optimal control problem of the energy management including NO$_\mathrm{x}^\mathrm{eo}$ emissions (from here on referred to as OCP) is to find the optimal power distribution for a predefined driving mission, such that the H$_2$ consumption is as low as possible, while the NO$_\mathrm{x}^\mathrm{eo}$ emissions are kept below an upper limit $\bar{m}^\mathrm{eo}_{\mathrm{NO}_\mathrm{x}}$. The OCP is written in time domain, i.e., the driving mission starts at the time instance $t_0$ and ends at time instance $t_f$. The vehicle has to be operated in charge-sustaining mode, i.e., the SoC must be the same at the end and at the start of the driving mission. Using the definitions for $\mathbf{x}$, $\mathbf{u}$, and equations $(\ref{eq:omega_w})$ -- $(\ref{eq:Voc})$ as constraints, the OCP is written as follows:
\begin{subequations}\label{eq:OCP}
	\begin{alignat}{2}
		\underset{\mathbf{u}}{\text{min}} \quad \quad \, & \int_{t_0}^{t_{\text{f}}}  \dot{m}_f \, dt \label{eq:OCP_cost_fcn}\\
		\text{   \qquad s.t.} \quad T_{\mathrm{req}_\mathrm{fd}} & = T_{e_\mathrm{req}} + T_{\mathrm{mot}_\mathrm{req}} \label{eq:OCP_Treq}\\
			\dot{\mathrm{SoC}} & = f(\mathbf{x}, \mathbf{u}) \label{eq:OCP_bat_state}\\
			\dot{m}^\mathrm{eo}_{\mathrm{NO}_\mathrm{x}} & = f(\mathbf{u}) \label{eq:OCP_NOx_state}\\
			\mathrm{SoC}(t_0) & = 0.7\\
			\mathrm{SoC}(t_f) & = 0.7 \label{eq:OCP_charge_sustainability}\\
			m^\mathrm{eo}_{\mathrm{NO}_\mathrm{x}}(t_0) & = 0\\
			m^\mathrm{eo}_{\mathrm{NO}_\mathrm{x}}(t_f) & \leq \bar{m}^\mathrm{eo}_{\mathrm{NO}_\mathrm{x}} \label{eq:OCP_NOx_terminalCondition}\\
			\mathbf{u} & \in \mathcal{U} \label{eq:OCP_inputSet}\\
			\mathbf{x} & \in \mathcal{X} \label{eq:OCP_stateSet}.
	\end{alignat}
\end{subequations}
Equation (\ref{eq:OCP_Treq}) is called the driveability constraint and ensures that the driver's torque request is fulfilled. The set of feasible inputs (\ref{eq:OCP_inputSet}) encompasses the constraints that are defined by (\ref{eq:omega_gen}), (\ref{eq:torqueSplit}),  (\ref{eq:generatorTorque}), as well as, the physical limitations of the drivetrain's components. The set of feasible state variables (\ref{eq:OCP_stateSet}) encompasses constant bounds on SoC $\in [0.3,\,0.9]$ and constant bounds on $m^\mathrm{eo}_{\mathrm{NO}_\mathrm{x}}\in [0, \bar{m}^\mathrm{eo}_{\mathrm{NO}_\mathrm{x}}]$.\\
\\
The OCP at hand is a nonlinear mixed-integer optimal control problem. Although, the dynamic programming algorithm (DP) could directly be used to solve the problem with optimality guarantees, the so-called  curse of dimensionality \cite{powell2007approximate} hinders its effective applicability. Therefore, a shooting algorithm is exploited to reduce the problem's dimensionality by eliminating the $m^\mathrm{eo}_{\mathrm{NO}_\mathrm{x}}$ state without jeopardizing the solution's global optimality. A slight adaption of the PMP-DP-algorithm that was proposed in \cite{uebel2017optimal} is used. A sketch of the idea is outlined in the following: Instead of solving the OCP (\ref{eq:OCP}) only for one upper bound $\bar{m}^\mathrm{eo}_{\mathrm{NO}_\mathrm{x}}$, the cost function (\ref{eq:OCP_cost_fcn}) is adjoined with an equivalent fuel consumption weight for the $\dot{m}^\mathrm{eo}_{\mathrm{NO}_\mathrm{x}}$ emissions, and this resulting adapted optimal control problem is solved multiple times for different weights. On the one hand, adjoining the $\dot{m}_{\mathrm{NO}_\mathrm{x}}$ dynamics allows to omit the constraint (\ref{eq:OCP_NOx_terminalCondition}), which can be leveraged to greatly decrease the computational burden, compared to solving the original OCP with the DP algorithm. On the other hand, this shooting method results in a Pareto front for the H$_2$-NO$_\mathrm{x}^\mathrm{eo}$ trade-off, which will be discussed in the following.

\section{Results: WLTC} \label{sec:Results}
\begin{figure}[h!]
    \centering
    \includegraphics[scale=0.425]{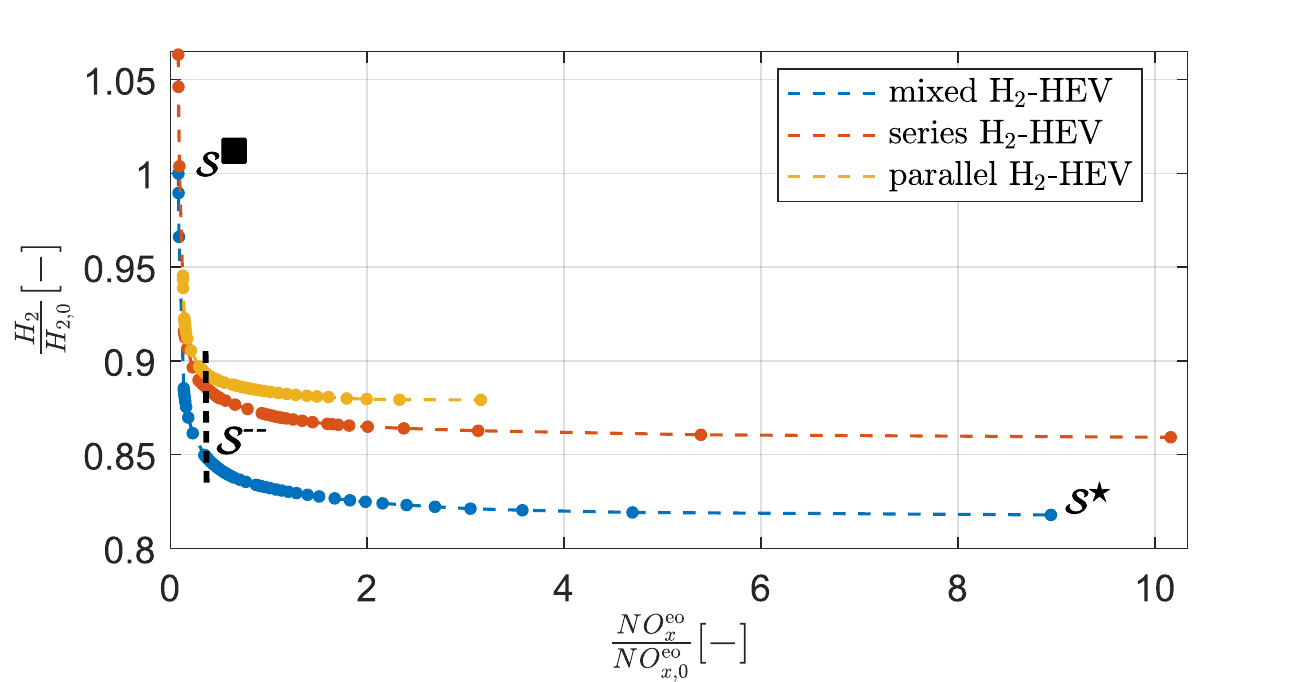}
    \caption{Achievable H$_2$-NO$_\mathrm{x}^{\mathrm{eo}}$ trade-off for the mixed H$_2$-HEV, the series H$_2$-HEV, and the parallel H$_2$-HEV. The results are normalized using H$_{2,0}$ and NO$_{\mathrm{x},0}^{\mathrm{eo}}$, both of which are obtained by simulations of the base H$_2$-vehicle. The black, dotted line represents an arbitrary accumulated NO$_{\mathrm{x},0}^{\mathrm{eo}}$ value and is used for comparison.}
    \label{fig:trade_off}
\end{figure}

The performance of the investigated propulsion architectures is analyzed using the results obtained with the adapted PMP-DP algorithm. Depending on the choice of the upper limit $\bar{m}^\mathrm{eo}_\mathrm{NO_\mathrm{x}}$ in (\ref{eq:OCP_NOx_terminalCondition}), different optimal EMS calibrations are realized for the same driving mission, which results in a Pareto front. All  Pareto fronts in this paper are normalized using H$_{2,0}$ and NO$_{\mathrm{x},0}^{\mathrm{eo}}$, which are the H$_2$ consumption and NO$_\mathrm{x}^{\mathrm{eo}}$ emissions of the base H$_2$-vehicle, respectively. As for the base H$_2$-vehicle, no torque split can be optimized, these values are fixed to a corresponding driving mission.\\
\\
Figure \ref{fig:trade_off} shows the trade-off between the H$_2$ consumption and the NO$_\mathrm{x}^{\mathrm{eo}}$ emissions for the WLTC. The different hybridized propulsion configurations are shown in yellow, red, and blue for the parallel, series, and mixed H$_2$-HEV, respectively. Depending on the torque distribution, different H$_2$ consumptions and NO$_\mathrm{x}^\mathrm{eo}$ emissions ensue. From here on, the H$_2$-optimal EMS calibration of the mixed H$_2$-HEV will be indicated by $\mathcal{S}^\star$ and is depicted by the point farthest to the right of the corresponding Pareto front. The NO$_\mathrm{x}^{\mathrm{eo}}$-optimal EMS calibration of the mixed H$_2$-HEV will be indicated by $\mathcal{S}^\blacksquare$ and is depicted by the point the farthest to the left of the corresponding Pareto front. Each point on the corresponding Pareto front represents a realizable EMS calibration which results in a unique combination of NO$_\mathrm{x}^{\mathrm{eo}}$ emissions and H$_2$ consumption, demonstrating that a hybridized powertrain architecture opens a large range of realizable trade-offs. To enable a comparative assessment of the individual vehicles, a specific level of accumulated NO$_\mathrm{x}^{\mathrm{eo}}$ emissions is established, and the corresponding hydrogen consumptions of each vehicle are then compared. This concept is graphically represented by the black dashed line. To facilitate below analysis, the intersection of the mixed H$_2$ HEV's Pareto front and black dashed line is the EMS calibration denoted by $\mathcal{S}^{--}$.
\subsection{Base vehicle vs mixed H$_2$-HEV}
\begin{figure}[h!]
    \centering
    \includegraphics[scale=0.35]{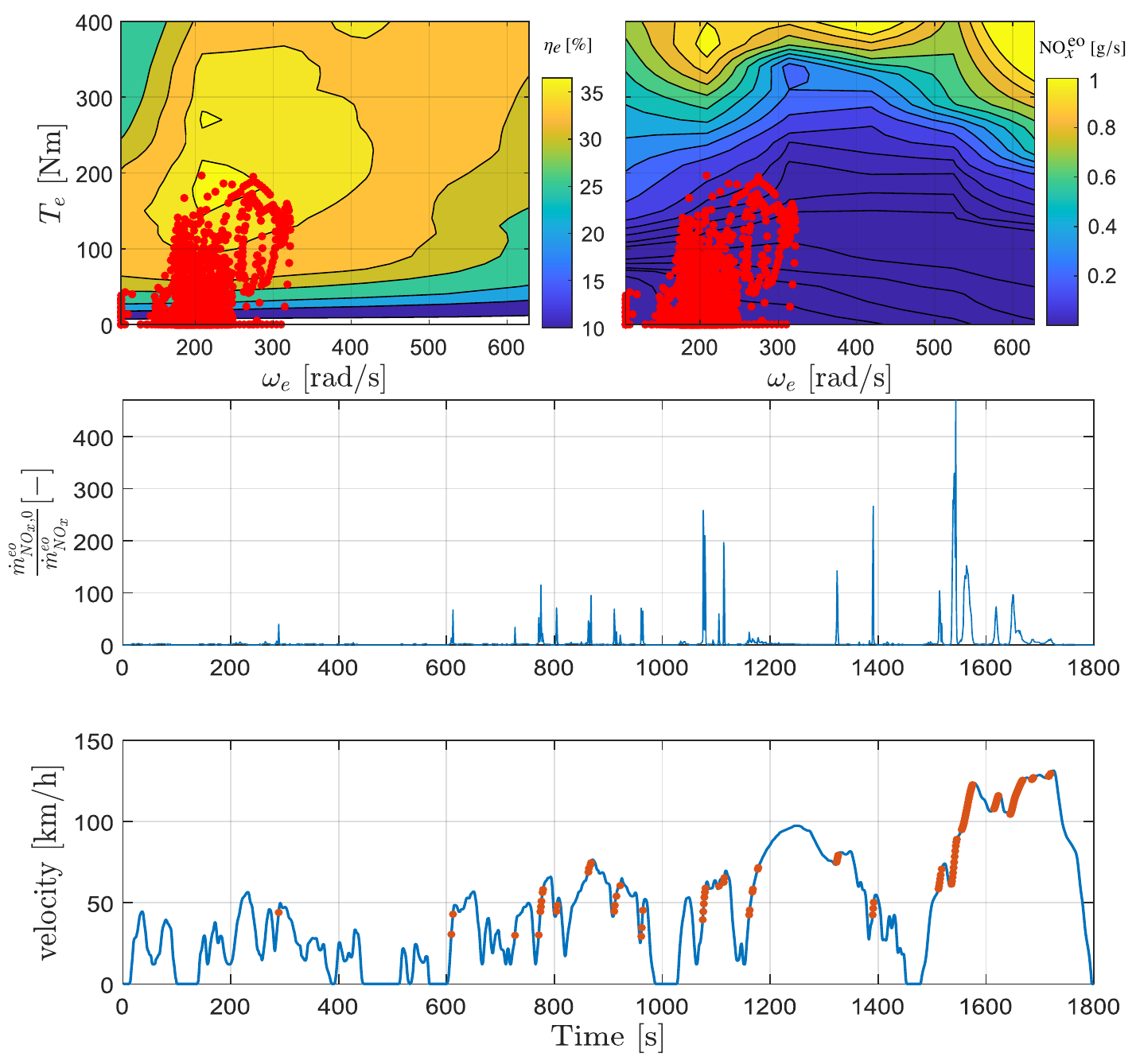}
    \caption{Base H$_2$-vehicle operating conditions on the WLTC. Top: Engine operating points superimposed on the engine efficiency map and the NO$_\mathrm{x}^{\mathrm{eo}}$ map. Middle: Time-resolved and normalized NO$_\mathrm{x,0}^{\mathrm{eo}}$ emissions. Bottom: Velocity profile of the WLTC with NO$_\mathrm{x,0}^{\mathrm{eo}}$ peaks marked by red dots.}
    \label{fig:H2ICE}
\end{figure}

The blue line depicts the Pareto front, which can be achieved using the mixed H$_2$-HEV. The EMS calibration, which emits the same NO$_\mathrm{x}^{\mathrm{eo}}$ emissions as the base H$_2$-vehicle, is located where its x-value equals to one. As a result, this simulation study reveals that for the same NO$_\mathrm{x}^\mathrm{eo}$ emissions, an H$_2$ reduction potential of 16.8\% is possible with the use of the mixed H$_2$-HEV. Even more remarkable is that the realizable NO$_\mathrm{x}^{\mathrm{eo}}$ emissions by the mixed H$_2$-HEV can be more than one order of magnitude lower than what is achieved by the base H$_2$-vehicle. This suggests that the base H$_2$-vehicle cannot be operated under ultra-lean combustion conditions throughout the entire drive mission.\\
\\
To investigate the base H$_2$-vehicle's NO$_\mathrm{x}^\mathrm{eo}$ emissions in detail, the operating conditions of its H$_2$ICE are analyzed below. Figure \ref{fig:H2ICE} shows in the upper plot all operating points of the base vehicle in the engine efficiency map and the NO$_\mathrm{x}^{\mathrm{eo}}$ map. Two things can be noted here: Firstly, many operating points lie in the low-load region, which is a well-known issue resulting in a low mean engine efficiency compared to hybrid powertrain configurations. Secondly, the NO$_\mathrm{x}^{\mathrm{eo}}$ map predicts a steep increase in NO$_\mathrm{x}^{\mathrm{eo}}$ production for increased engine torques. Although, most of the engine operating points are located in a region where very low  NO$_\mathrm{x}^{\mathrm{eo}}$ emissions occur, some operating points of the base H$_2$-vehicle lie critically close to the steep region.\\
To investigate these critical points in detail, the middle plot shows a time-resolved graph of the NO$_{x,0}^{\mathrm{eo}}$ emitted by the base H$_2$-vehicle normalized by the NO$_\mathrm{x}^{\mathrm{eo}}$ emitted by $\mathcal{S}^{--}$ of the mixed H$_2$-HEV. It shows that, most of the time, both vehicles have NO$_\mathrm{x}^\mathrm{eo}$ emissions, which have the same order of magnitude. However, the spikes in this signal reveal where the base H$_2$-vehicle emits substantially more NO$_\mathrm{x}^{\mathrm{eo}}$.\\
The lowest plot shows the velocity profile of the investigated WLTC. The driving scenarios leading to the NO$_\mathrm{x}^{\mathrm{eo}}$-spikes are highlighted by red dots. They occur at high vehicle speed, i.e., highway driving, or when accelerating with more than 1 m/s$^2$ during urban driving. Since such driving patterns are common in daily road traffic it can be concluded that the base H$_2$-vehicle cannot be operated under extremely low NO$_\mathrm{x}^{\mathrm{eo}}$ emissions conditions in normal driving. 

\subsection{Hybridized drivetrains}
As the base H$_2$-vehicle cannot reliably ensure extremely low NO$_\mathrm{x,0}^{\mathrm{eo}}$ emissions, an investigation is conducted below about how the mixed H$_2$-HEV is able to reach lower NO$_\mathrm{x}^{\mathrm{eo}}$ emissions. Comparing the red and the yellow lines in Figure \ref{fig:trade_off} shows that the H$_2$ saving potential of the series H$_2$-HEV and the parallel H$_2$-HEV are comparable on the WLTC. The EMS calibrations, which are indicated by the intersections of the Pareto fronts with the black, dotted line reveal that the series H$_2$-HEV is able to obtain a 0.8\% lower H$_2$ consumption compared to the parallel H$_2$-HEV. With the mixed H$_2$-HEV architecture however, the H$_2$ saving potential is 5\%, compared to the parallel H$_2$-HEV. To understand how the mixed H$_2$-HEV's EMS is able to reduce the accumulated NO$_\mathrm{x}^{\mathrm{eo}}$ emissions, the solutions $\mathcal{S}^\star$ and $\mathcal{S}^{--}$ of this drivetrain architecture are compared below.
\begin{figure}[h!]
    \centering
    \includegraphics[scale=0.425]{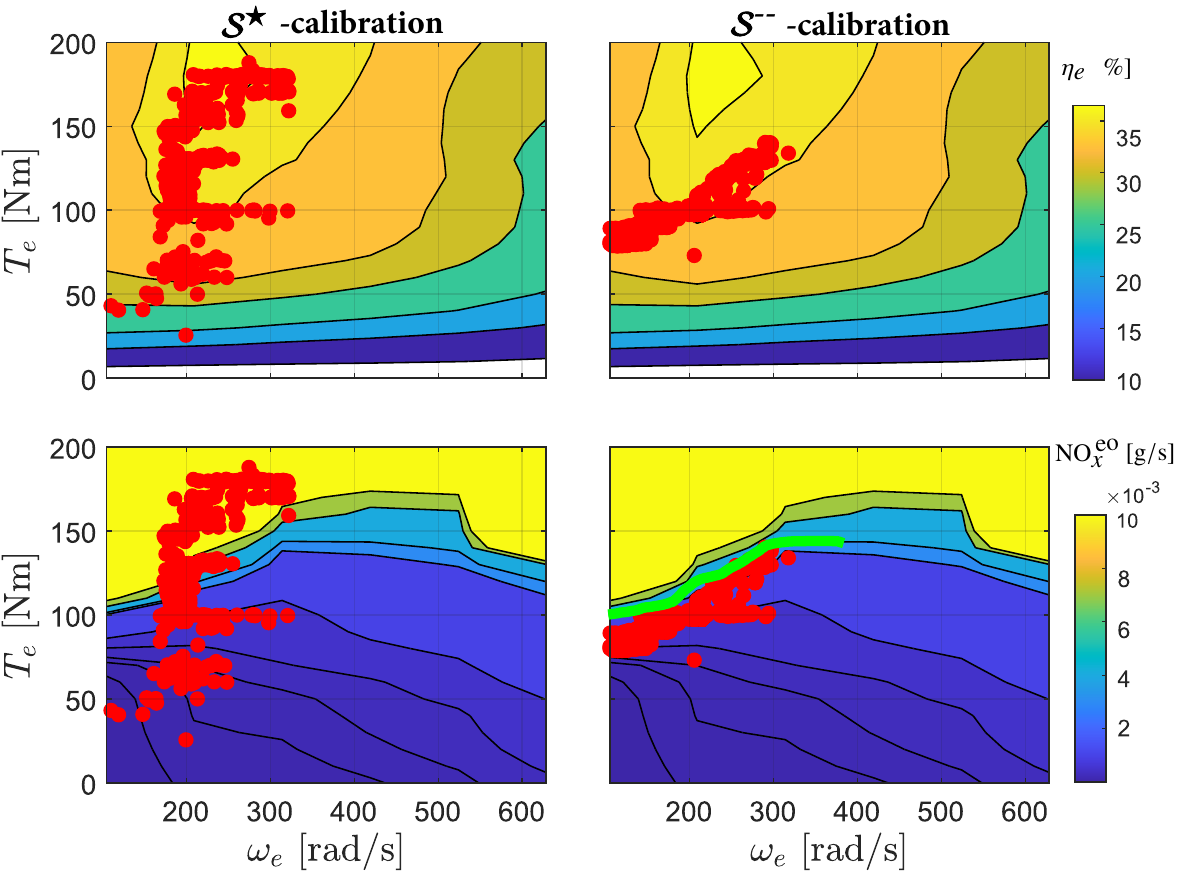}
    \caption{Left: Engine operating points superimposed on the engine efficiency map and the NO$_\mathrm{x}^{\mathrm{eo}}$ map resulting of the EMS calibration $\mathcal{S}^\star$. Right: Engine operating points superimposed on the engine efficiency map and the NO$_\mathrm{x}^{\mathrm{eo}}$ map resulting of the EMS calibration $\mathcal{S}^{--}$.}
    \label{fig:mixedHEV}
\end{figure}
\\
Figure \ref{fig:mixedHEV} depicts the engine operating points of these two solutions, superimposed on the engine efficiency map and the NO$_\mathrm{x}^{\mathrm{eo}}$ map. Two things can be noticed here: Firstly, it is optimal in both scenarios to omit low-load engine operating points, i.e., $T_e \lessapprox 45$ Nm. This is a well-known result and usually one of the main drivers to reduce the overall fuel consumption of any HEV equipped with a combustion engine. Secondly, the optimal EMS calibration $\mathcal{S}^{--}$ is achieved by omitting too high-load engine operating points. The maximum engine torque does not surpass the NO$_\mathrm{x}^{\mathrm{eo}}$-isoline, which is highlighted by the green line. Above this line, the emitted NO$_\mathrm{x}^{\mathrm{eo}}$ increase rapidly, as ultra-lean operation of the H$_2$ICE is no longer possible.
\section{Different Driving Missions} \label{sec:different_driving_missions}
So far, it was shown that the mixed H$_2$-HEV has superior performance compared to the other H$_2$-HEVs and the base H$_2$-vehicle. In the following, to further highlight the potential for reducing the NO$_\mathrm{x}^{\mathrm{eo}}$ of this drivetrain structure, a specific EMS calibration is selected, which achieves 90\% less NO$_\mathrm{x}^{\mathrm{eo}}$ than the base H$_2$ vehicle. It is denoted by $\mathcal{S}^\blacktriangle$.\\
\\
The remainder of this publication is used to investigate whether the potential of the mixed H$_2$-HEV to reduce the NO$_\mathrm{x}^{\mathrm{eo}}$ emissions can be generalized to other driving missions. Consequently, four additional driving missions are introduced and their corresponding Pareto fronts are presented. The first two represent more realistic driving scenarios. The third and the fourth driving mission are used to show the limitations of the mixed H$_2$-HEV powertrain architecture to reduce NO$_\mathrm{x}^{\mathrm{eo}}$.\\
\\
Finally, most state-of-the-art EMS algorithms (e.g., \cite{beccari2022use}, \cite{kyjovsky2023drive}) only pay attention to minimizing the H$_2$ consumption resulting (in the best case) in $\mathcal{S}^\star$. As in this work, the possibility to decrease the NO$_\mathrm{x}^{\mathrm{eo}}$ at the expense of an increased H$_2$ consumption is discussed, three additional key parameters are introduced below. On the one hand, they serve the purpose of facilitating a contrast between different realizations of the H$_2$-NO$_\mathrm{x}^{\mathrm{eo}}$ trade-off, and on the other hand, they emphasize the advantages of the hybridization of the base H$_2$-vehicle. In particular, the EMS calibration $\mathcal{S}^\blacktriangle$ is investigated, as it provides a substantial decrease in NO$_\mathrm{x}^{\mathrm{eo}}$ at a marginally increased H$_2$ consumption. For convenience, and to enable a comparison among the individual driving missions, all key parameters are summarized in Table \ref{tab:performanceSummary}.\\
\\
\textit{Key performance parameters:}\\
Firstly, if not the H$_2$-optimal EMS calibration is chosen, but instead the EMS calibration $\mathcal{S}^\blacktriangle$, the following increase in H$_2$ consumption occurs:
\begin{equation}
	\Delta \mathrm{H}_2^\mathrm{add} = 1 - \frac{\mathrm{H}_2^{\mathcal{S}^\star}}{\mathrm{H}_2^{\mathcal{S}^\blacktriangle}}.
\end{equation}
Secondly, the highest possible NO$_\mathrm{x}^{\mathrm{eo}}$ emissions reduction of the mixed H$_2$-HEV compared to the base H$_2$-vehicle calculates as:
\begin{equation}
		f_\mathrm{x}^{\mathrm{eo,}\mathcal{S}^\blacksquare} = \frac{\mathrm{NO}_\mathrm{x}^{\mathrm{eo,}\mathcal{S}^\blacksquare}}{\mathrm{NO}_\mathrm{x,0}^{\mathrm{eo}}}.
\end{equation}
It is achieved by the EMS calibration $\mathcal{S}^\blacksquare$ and it describes the mixed H$_2$-HEV's limit to reduce NO$_\mathrm{x}^{\mathrm{eo}}$.\\
\\
Thirdly, the last key parameter describes how much H$_2$ the mixed H$_2$-HEV saves, while aiming for a reduction of 90\% NO$_\mathrm{x}^{\mathrm{eo}}$ emissions, compared to the base H$_2$-vehicle:
\begin{equation}
		\Delta \mathrm{H}_2 = 1 - \frac{\mathrm{H}_2^{\mathcal{S}^\blacktriangle}}{\mathrm{H}_{2,0}}.
\end{equation}
It is mainly used to showcase how powerful the hybridization of a vehicle with an H$_2$ICE can be.
\subsection{Realistic Driving Scenarios}\label{subsec:realistic_scenarios}
\begin{figure*}[h!]
    \includegraphics[width=\textwidth]{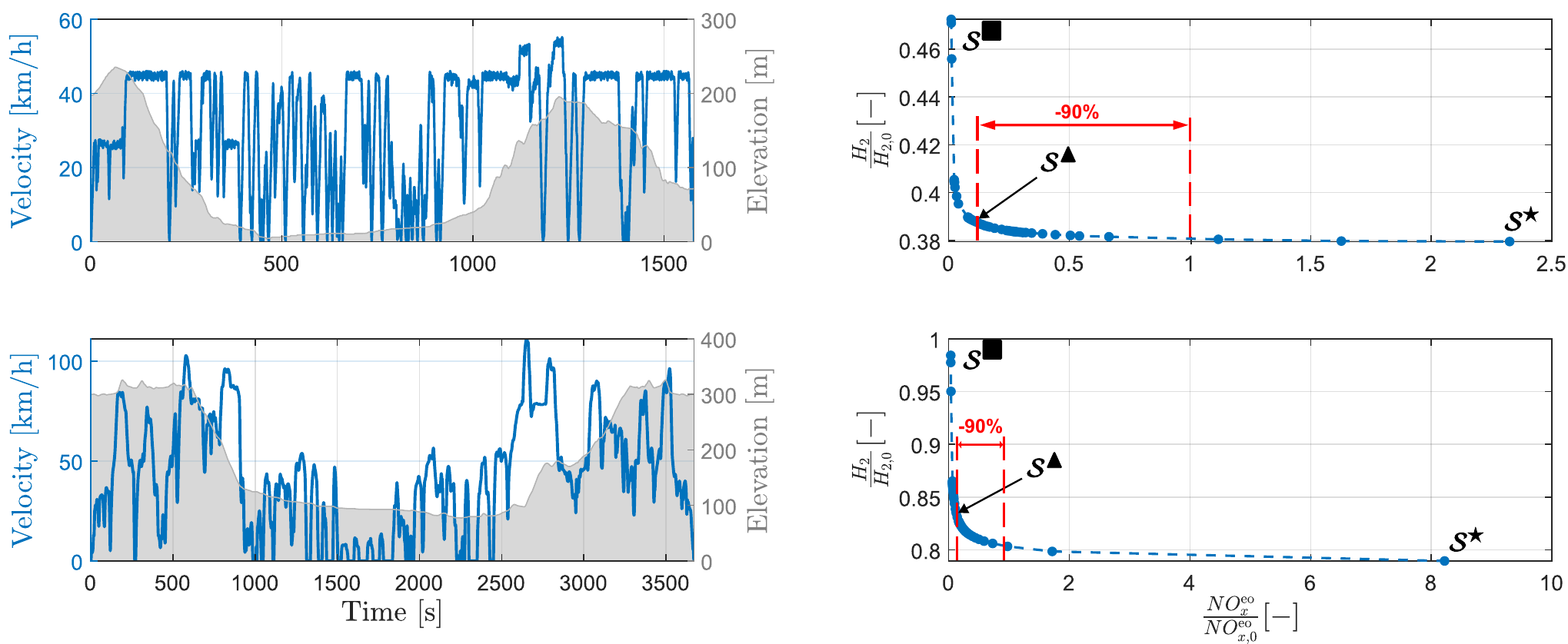}
    \caption{Left: Driving missions: "\textit{urban cycle}" and "\textit{real driving cycle}". Right: H$_2$-NO$_\mathrm{x}^{\mathrm{eo}}$ Pareto-optimal operations of the mixed H$_2$-HEV for the corresponding driving missions. The results are normalized using H$_{2,0}$ and NO$_{x,0}^{\mathrm{eo}}$ of the base H$_2$-vehicle. Three EMS calibrations, i.e., $\mathcal{S}^\star$, $ \mathcal{S}^\blacktriangle$, and $\mathcal{S}^\blacksquare$ are highlighted for the mixed H$_2$-HEV and both driving cycles.}
    \label{fig:RDE_ZH_cycle}
\end{figure*}
On the left hand side of Figure \ref{fig:RDE_ZH_cycle}, two additional realistic driving scenarios are depicted. The first driving mission is extracted from the open-source software SUMO \cite{behrisch2011sumo} and is referred to as \textit{urban cycle}. The second driving mission is based on recorded sensor data of a real vehicle and features a mix of urban driving, rural driving, and highway driving. It is referred to as \textit{real driving cycle}. Both missions feature an extensive elevation profile.\\
\\
On the right hand side of Figure \ref{fig:RDE_ZH_cycle}, the corresponding H$_2$-NO$_\mathrm{x}^{\mathrm{eo}}$ Pareto fronts are depicted for both driving missions. For both driving missions, there exist EMS calibrations $\mathcal{S}^\blacktriangle$ achieving the intended significant NO$_\mathrm{x}^{\mathrm{eo}}$ reductions. The factor $\Delta \mathrm{H}_2^\mathrm{add}$ shows that these large reductions in NO$_\mathrm{x}^{\mathrm{eo}}$ come at an increased hydrogen consumption of only 2.2\% (\textit{urban cycle}), or 5.8\% (\textit{real driving cycle}). The factor $\Delta \mathrm{H}_2$ shows that it is possible to reduce the base H$_2$-vehicle's NO$_\mathrm{x}^\mathrm{eo}$ emissions by 90\%, while still saving 61.2\% (\textit{urban cycle}), or 16.1\% (\textit{real driving cycle}), respectively. Finally, an even lower NO$_\mathrm{x}^{\mathrm{eo}}$ emission level can be achieved, i.e., \mbox{$f_\mathrm{x}^{\mathrm{eo,}\mathcal{S}^\blacksquare}=$ 99\%} and \mbox{$f_\mathrm{x}^{\mathrm{eo,}\mathcal{S}^\blacksquare}=$ 96.6\%} on the \textit{urban cycle}, and the \textit{real driving cycle}, respectively.
\subsection{Worst-case driving scenarios}\label{sec:worst_case}
\begin{figure*}[h!]
	\centering
	\includegraphics[width=\textwidth]{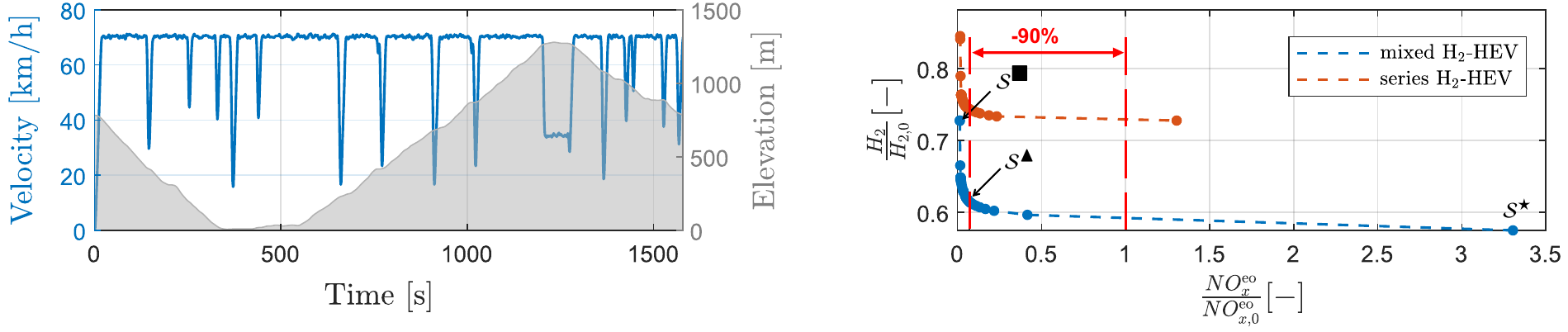}
	\caption{Left: Driving missions: "\textit{mountain cycle}". Right: H$_2$-NO$_\mathrm{x}^\mathrm{eo}$ Pareto-optimal operations of the mixed H$_2$-HEV for the "\textit{mountain cycle}". The results are normalized using H$_{2,0}$ and NO$_{x,0}^{\mathrm{eo}}$ of the base H$_2$-vehicle. Three EMS calibrations, i.e., $\mathcal{S}^\star$, $ \mathcal{S}^\blacktriangle$, and $\mathcal{S}^\blacksquare$ are highlighted for the mixed H$_2$-HEV.}
	\label{fig:mountain_cycle}
\end{figure*}

\begin{figure}[h!]
    \centering
    \includegraphics[scale=0.45]{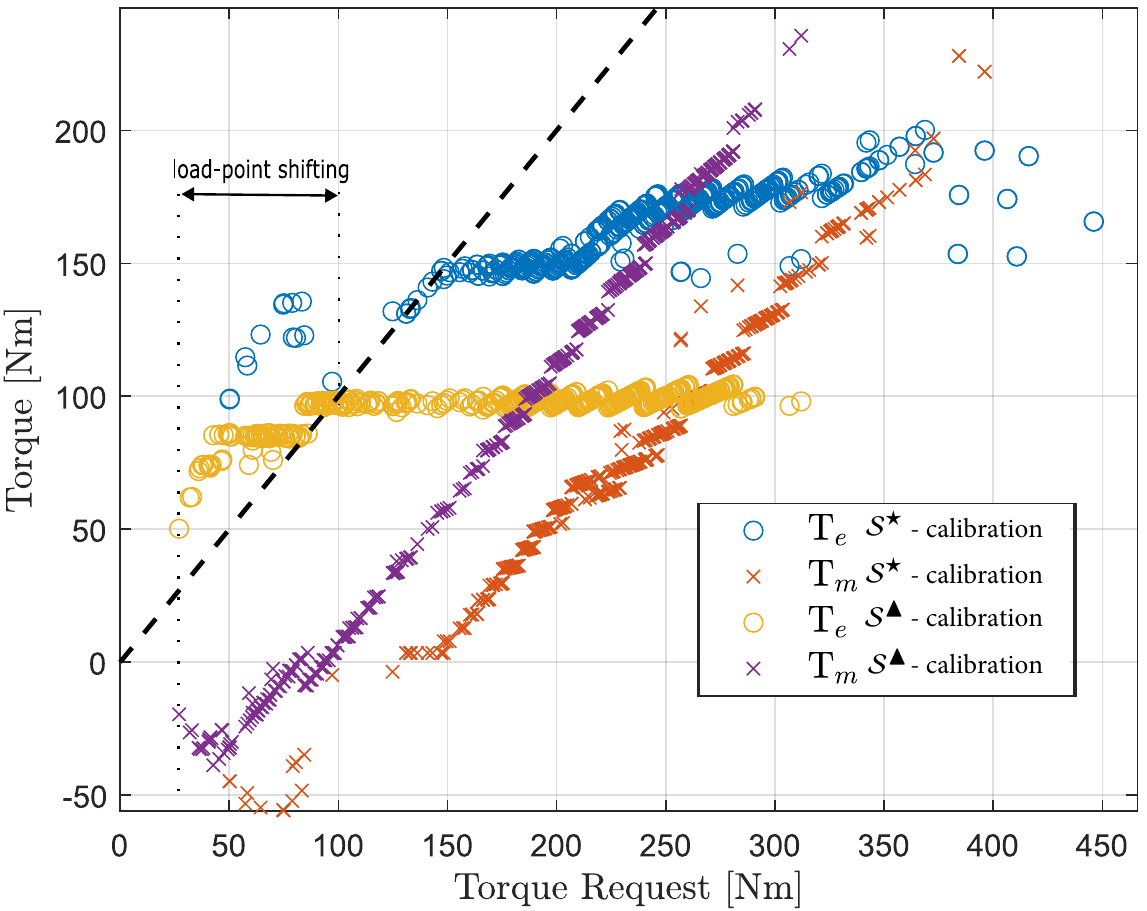}
    \caption{Distribution of the torque request of two optimal EMS calibrations (i.e., $\mathcal{S}^\blacktriangle$, $\mathcal{S}^\blacksquare$) of the mixed H$_2$-HEV on the \textit{mountain cycle}. Only the parallel mode is displayed. The black dashed line represents the 45$^\circ$-line.}
    \label{fig:TreqSplit}
\end{figure}

\begin{figure}[h!]
    \centering
    \includegraphics[scale=0.5]{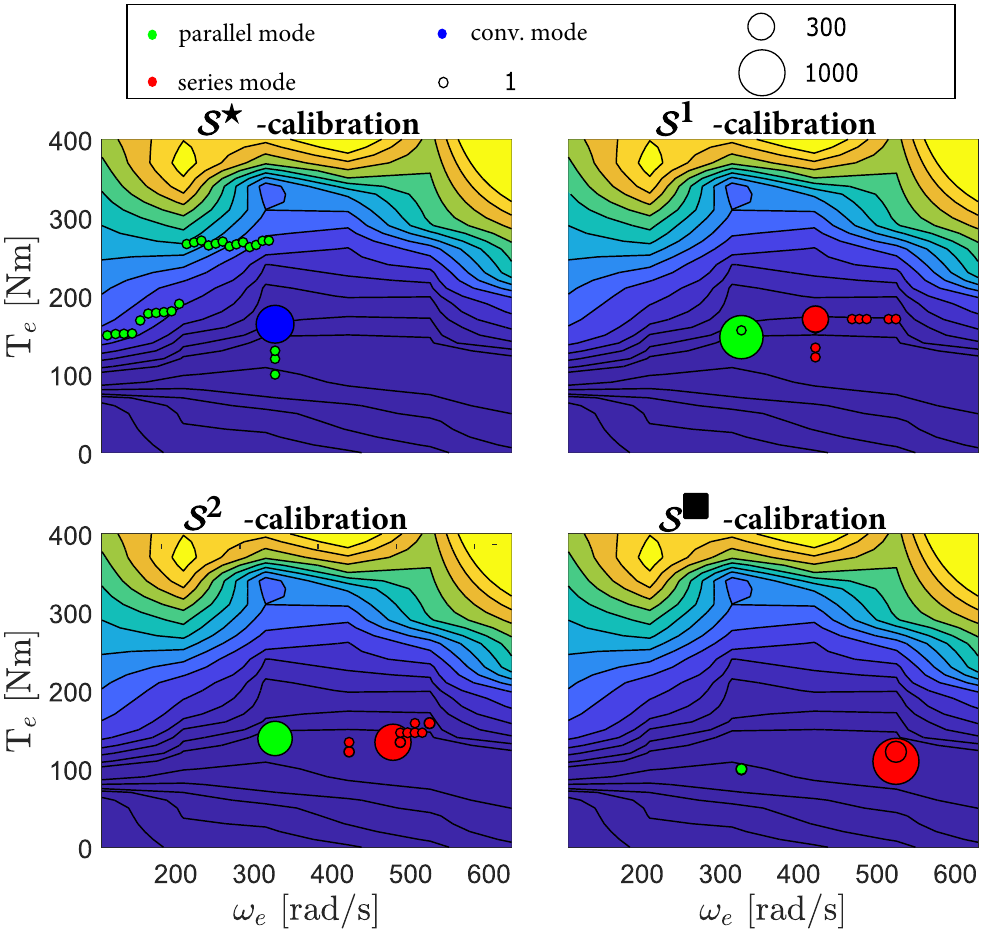}
    \caption{Engine operating points on the \textit{highway cycle}. The size of the dots represents how often that the corresponding operating point has been chosen.}
    \label{fig:Highway_NOxCompliant}
\end{figure}

The analysis of the Pareto fronts obtained for the \textit{urban cycle} and the \textit{real driving cycle} show that for driving patterns that are likely to occur during day-to-day driving, the mixed H$_2$-HEV provides a large range of H$2$-NO$_\mathrm{x}^{\mathrm{eo}}$ trade-offs to choose from. This section now focuses on extreme driving scenarios, which are chosen to show the limitations of the mixed H$_2$-HEV. The main tool to decrease the NO$_\mathrm{x}^\mathrm{eo}$ emissions via hybridization of the propulsion architecture bases on omitting too high engine-load operating points. So far, the investigated driving missions all include a broad range of different torque requests, and, for these missions, the results obtained by DP show that the mixed H$_2$-HEV is a promising topology to enable a large range of realizable H$_2$-NO$_\mathrm{x}^\mathrm{eo}$ trade-offs. Therefore, two additional driving missions are presented in the following, which mainly consist of high torque requests.\\
\\
\textit{Mountain Cycle}:\\
The left hand side of Figure \ref{fig:mountain_cycle} depicts the \textit{mountain cycle}, which is inspired by a mountain pass in the Swiss alps, and represents a trip including a continuous climb of over \mbox{1270 m}. It is assumed that no other road users are present, i.e., other than in the hairpin turns, the vehicle speed is constant at \mbox{70 km/h}.\\
The right hand side of Figure \ref{fig:mountain_cycle} depicts in blue the achievable Pareto-optimal H$_2$-NO$_\mathrm{x}^\mathrm{eo}$ trade-off for the mixed H$_2$-HEV on the \textit{mountain cycle}. To  highlight the importance of the parallel mode for this kind of driving missions below, the series H$_2$-HEV's Pareto front is shown in red. There exists an EMS calibration $\mathcal{S}^\blacktriangle$, which achieves the intended significant NO$_\mathrm{x}^\mathrm{eo}$ reductions. It results in an increase of hydrogen consumption of $\Delta H_2^\mathrm{add} = 5.9\%$. The factor $\Delta \mathrm{H}_2$ shows that it is possible to reduce the base H$_2$-vehicle's emissions by 90\%, while saving 38.9\% H$_2$. Finally, an even lower NO$_\mathrm{x}^\mathrm{eo}$ emission level can be achieved, i.e., \mbox{$f_\mathrm{x}^{\mathrm{eo,}\mathcal{S}^\blacksquare}=$ 98.6\%}, showing that despite the considerable power demand required for ascending the elevation, the mixed H$_2$-HEV is able to decrease the NO$_\mathrm{x}^\mathrm{eo}$ emissions with a proper EMS calibration.\\
\\
Comparing the blue and red Pareto fronts shows that the mixed H$_2$-HEV is able to utilize the parallel mode to effectively decrease the H$_2$ consumption. How the parallel mode is used, is outlined in the following: Figure \ref{fig:TreqSplit} displays the optimal distribution of the torque request between the engine and the motor, evaluated at the differential. The torque distribution is only shown for the parallel mode of the two Pareto-optimal EMS calibrations $\mathcal{S}^\blacktriangle$ and $\mathcal{S}^\star$. First of all, the optimal engine usage (depicted by the circles) is analyzed for both EMS calibrations: For torque requests below \mbox{100 Nm} it is optimal to use load-point shifting (,i.e., increasing the produced engine power above what is required at the wheel to store excess power via the motor in the battery) to additionally recharge the battery. This is depicted by $T_e$ above the dashed 45$^\circ$-line. This  allows to simultaneously operate the H$_2$ICE in a higher efficiency operating point (without getting into the high-NO$_\mathrm{x}^\mathrm{eo}$ region) and to further charge the battery with electric energy that can be spent later on. The total positive electric energy throughput in EV mode is decreased by roughly 65\% (from 4.1 MJ in $\mathcal{S}^\star$ to 2.6 MJ in $\mathcal{S}^\blacktriangle$). This shows that, over the entire driving mission, the usage of EV mode is decreased, and the battery electric energy is increasingly used in parallel mode. While driving in parallel mode, the electric energy is used to relieve the engine of too high load-points that occur during uphill driving. Focusing on the blue circles, the H$_2$-optimal EMS calibration is obtained by respecting a (slightly scattered,) preferred maximum engine torque, which is achieved by additionally boosting with the electric motor (indicated by the red crosses). The electric motor is used such that the H$_2$ICE is run in its highest speed-dependent fuel-efficiency operating points (compare to Figure \ref{fig:fuelNOx}).\\
\\
The yellow circles and purple crosses represent the engine torque and the motor torque for the EMS calibration $\mathcal{S}^\blacktriangle$. Again, there exists an upper engine torque limit at around \mbox{100 Nm}, until which the entire torque request is supplied solely by the engine, and beyond which any additional torque request is supplied by boosting with the motor. However, here this bound is more distinct and lower than in $\mathcal{S}^\star$, revealing that the electric motor is used to prevent high engine torques.\\
\\
Comparing the blue and red Pareto fronts in Figure \ref{fig:mountain_cycle}, the importance of the parallel mode for this driving missions is demonstrated. Although, extremely low NO$_\mathrm{x}^{\mathrm{eo}}$ emissions can be achieved by the series H$_2$-HEV, these EMS calibrations can come at a relatively high H$_2$ consumption, when compared to the mixed H$_2$-HEV. In conclusion, electric boosting in parallel mode is utilized by the mixed H$_2$-HEV to operate the H$_2$ICE under extremely low NO$_\mathrm{x}^\mathrm{eo}$ conditions, while still keeping the H$_2$ consumption at a low level. However, this torque distribution strategy is only viable, if the driving missions offers enough potential for circulating energy, i.e., energy that is stored temporarily in the vehicle body as kinetic and potential energy and used to recharge the battery.\\
\\
\textit{Highway Cycle}:\\
\begin{figure*}[h!]
	\centering
	\includegraphics[width=\textwidth]{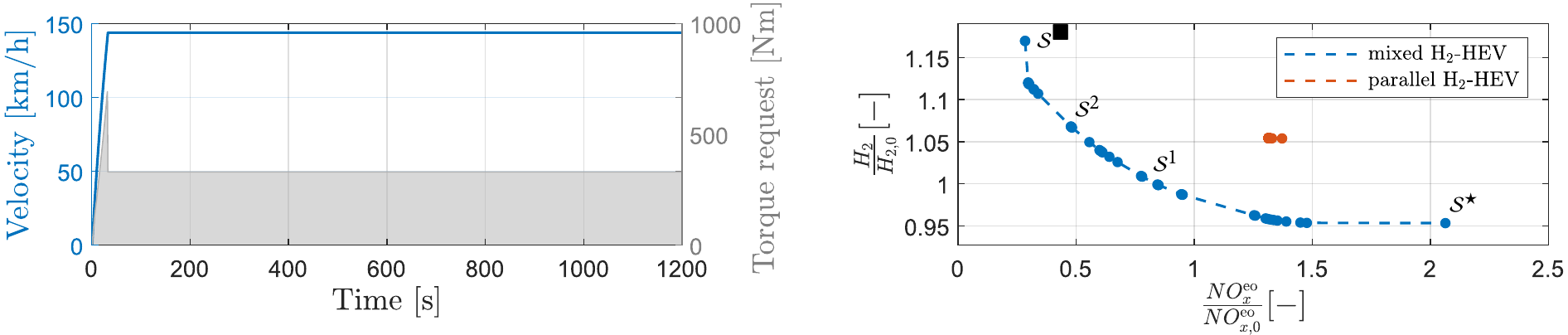}
	\caption{Left: Driving missions: "\textit{highway cycle}". Right: H$_2$-NO$_\mathrm{x}^\mathrm{eo}$ Pareto-optimal operations of the mixed H$_2$-HEV for the "\textit{highway cycle}". The results are normalized using H$_{2,0}$ and NO$_{x,0}^{\mathrm{eo}}$ of the base H$_2$-vehicle. Four EMS calibrations, i.e., $\mathcal{S}^\star$, $ \mathcal{S}^1$, $ \mathcal{S}^2$, and $\mathcal{S}^\blacksquare$ are highlighted for the mixed H$_2$-HEV.}
	\label{fig:highway_cycle}
\end{figure*}
As was shown for the \textit{mountain cycle}, circulating energy is leveraged by the mixed H$_2$-HEV to achieve a broad range of H$_2$-NO$_\mathrm{x}^\mathrm{eo}$ trade-offs. The following cycle is constructed such that it offers no circulating energy at all. The left hand side of Figure \ref{fig:highway_cycle} depicts the \textit{highway cycle}. Similar to the \textit{mountain cycle}, high torque requests occur in this scenario. The difference is that the \textit{highway cycle} is an artificial scenario, which is constructed to solely consist of high, positive torque requests:
After a short acceleration phase, the vehicle reaches a constant speed of \mbox{142 km/h}, which requires a torque request of \mbox{372 Nm} at the wheel to maintain the speed level for the rest of the driving mission. In parallel mode, this requirement translates to an engine speed of \mbox{$\omega_{e}$ = 352 rad/s}. If the engine was to provide the full torque request on its own, it would need to provide a torque of \mbox{$T_e=$ 153 Nm}, which results in a good fuel efficiency, but cannot be sustained under ultra-lean combustion conditions and therefore results in high NO$_\mathrm{x}^\mathrm{eo}$ emissions. Decreasing the engine load via electric boosting and thereby lowering the NO$_\mathrm{x}^\mathrm{eo}$ is infeasible: On the one hand, the EMS is required to satisfy the charge-sustainability constraint (\ref{eq:OCP_charge_sustainability}) over the entire driving mission. On the other hand, this driving mission offers no opportunities to recuperate braking energy or to use load point shifting, as the goal is to reduce the engine load.\\
The right hand side of Figure \ref{fig:highway_cycle} depicts the achievable Pareto-optimal H$_2$-NO$_\mathrm{x}^\mathrm{eo}$ trade-offs for the the parallel H$_2$-HEV and the mixed H$_2$-HEV in red and blue, respectively. Most notably is that the EMS calibration $\mathcal{S}^\blacktriangle$ cannot be reached, given that the reduction of NO$_\mathrm{x}^\mathrm{eo}$ emissions to that extent is unattainable. The lowest possible cumulated NO$_\mathrm{x}^\mathrm{eo}$ emissions are $f_\mathrm{x}^{\mathrm{eo,}\mathcal{S}^\blacksquare}$ = 71.6\% lower than what is achieved by the base H$_2$-vehicle. But, this reduction of NO$_\mathrm{x}^\mathrm{eo}$ comes at the cost of a greatly increased H$_2$ consumption, which is 17\% higher than the base H$_2$-vehicle, and 18.5\% higher than the H$_2$-optimal EMS calibration $\mathcal{S}^\star$. This significant rise in H$_2$ consumption is primarily due to the fact that the base H$_2$-vehicle can almost exclusively operate its H$_2$ICE at a highly efficient operating point, meaning that a hybridized drivetrain can not increase its mean efficiency by any significant amount. However, the mixed H$_2$-HEV is still able to offer great flexibility to adjust the H$_2$-NO$_\mathrm{x}^\mathrm{eo}$ trade-off, which heavily depends on the usage of the series mode.\\
\\
Figure \ref{fig:Highway_NOxCompliant} provides further insight into how four different optimal EMS calibrations, i.e., $\mathcal{S}^\star$, $ \mathcal{S}^1$, $\mathcal{S}^2$, and $\mathcal{S}^\blacksquare$, use the series mode to achieve the corresponding values for the cumulated NO$_\mathrm{x}^\mathrm{eo}$ emissions. For each EMS calibration, the complete set of engine operating points is shown in the NO$_\mathrm{x}^\mathrm{eo}$-emissions map. Following the graphs from the top left towards the bottom right, the individual strategies depicted reduce the cumulated NO$_\mathrm{x}^\mathrm{eo}$ step by step. Green circles indicate parallel mode with additional support of the electric motor, blue circles indicate parallel mode where only the H$_2$ICE produces power (comparable to a conventional vehicle), and red circles indicate series mode. The size of the circles indicates the amount of times that the corresponding operating point is chosen. For the H$_2$-optimal EMS calibration $\mathcal{S}^\star$, most of the cycle is driven solely using the H$_2$ICE. The parallel mode is used at the beginning to assist in the acceleration of the vehicle with the electric motor, before the H$_2$ICE takes over shortly before reaching 142 km/h.\\
When gradually decreasing the cumulated NO$_\mathrm{x}^\mathrm{eo}$, three changes in the optimal EMS calibration can be identified. Firstly, pure conventional driving is substituted by an electrically assisted use of the H$_2$ICE, which is indicated by replacing the blue circle with a green one. The electric energy needed to ensure charge-sustainability is generated via the series mode. Secondly, when transitioning from $\mathcal{S}^1$ to $\mathcal{S}^2$, a shift from parallel mode to series mode is observed, which is indicated by the growing size of the red circles and a shrinking size of the green circle. This ultimately leads to solely operating the vehicle in series mode. Thirdly, in series mode, the H$_2$ICE operating points are shifted to ever higher engine speeds the lower the cumulated NO$_\mathrm{x}^\mathrm{eo}$ emissions get. The reason for this transition is given by the fact that it allows to sustain the engine power output, while decreasing the engine torque. This enables the engine-generator unit to deliver electric power to the motor or the battery, without increasing (or even decreasing) the NO$_\mathrm{x}^\mathrm{eo}$ emissions. However, this shift comes at the price of an increased H$_2$ consumption, as the engine efficiency drops above approximately 230 rad/s. In conclusion, the series mode is utilized by the mixed H$_2$-HEV to achieve extremely low cumulated NO$_\mathrm{x}^\mathrm{eo}$ emissions, even if not enough circulated energy can be recuperated. However, there is a limit to the maximum torque request that can be provided under ultra-lean combustion conditions. At some point, the torque request is too large and the engine-generator unit cannot shift to ever higher engine speeds. As a result, $T_e$ has to be increased, which directly leads to increased NO$_\mathrm{x}^\mathrm{eo}$ emissions.
\begin{table}[h!]
\vspace{-0.2cm}
\centering
\caption{Summary of key performance indicators of the mixed H$_2$-HEV for various investigated driving missions. As $\mathcal{S}^\blacktriangle$ cannot be realized on the \textit{highway cycle}, the metrics "$\Delta H_2^\mathrm{add}$" and "$\Delta H_2$" are calculated with respect to $\mathcal{S}^\blacksquare$, which is denoted with the superscript "$^\blacksquare$" in the corresponding columns.}
\label{tab:performanceSummary}
\begin{tabular}{l| c| c| c}
 & $\Delta \mathrm{H}_2^\mathrm{add}$ & $\Delta \mathrm{H}_2$ & $f_\mathrm{x}^\mathrm{eo,\mathcal{S}^\blacksquare}$\\\hhline{-|-|-|-}
\textit{Urban cycle} & 2.2\% & 61.2\% & 99\% \\\hhline{-|-|-|-} %
\textit{Real driving cycle} & 5.8\% & 16.1\% & 96.6\% \\\hhline{-|-|-|-} %
\textit{Mountain cycle} & 5.9\% & 38.9\% & 98.6\%\\\hhline{-|-|-|-} %
\textit{Highway cycle} & 18.5\%$^\blacksquare$ & -17\%$^\blacksquare$ & 71.6\%\\\hhline{-|-|-|-} %
\end{tabular}
\end{table}
\section{Conclusion} \label{sec:Conclusion}
To the authors' best knowledge, this paper provides the first full potential analysis of electrically hybridized vehicles equipped with an H$_2$ICE. A wide flexibility for choosing a Pareto-optimal trade-off between H$_2$ consumption and NO$_\mathrm{x}^\mathrm{eo}$ emissions can be achieved, solely by changing the distribution of the torque request between the vehicle's power sources. Overall, the performance of three different hybridized powertrain solutions, i.e., a series H$_2$-HEV, a parallel H$_2$-HEV, and a mixed H$_2$-HEV, are compared to a base H$_2$-vehicle on the WLTC. The resulting Pareto-optimal H$_2$-NO$_\mathrm{x}^\mathrm{eo}$ trade-offs show two things: Firstly, all investigated H$_2$-HEVs greatly outperform the base H$_2$-vehicle. Secondly, the mixed H$_2$-HEV is the superior H$_2$-HEV architecture in both regards: H$_2$ consumption and NO$_\mathrm{x}^\mathrm{eo}$ emissions.\\
\\
To solidify the findings obtained on the WLTC, the performance potential of the mixed H$_2$-HEV is further investigated on a variety of different realistic driving missions. The mixed H$_2$-HEV effectively leverages both the parallel mode and the series mode to lower the H$_2$ consumption and reach extremely low NO$_\mathrm{x}^\mathrm{eo}$ emissions, respectively:
To reduce the H$_2$ consumption by increasing the drivetrain's overall efficiency, the parallel mode is used. To achieve extremely low NO$_\mathrm{x}^\mathrm{eo}$ emissions by decoupling the H$_2$ICE, which allows to run it in its most suitable operating point, the series mode is used.
For a broad range of driving missions, the mixed H$_2$-HEV is able to decrease the engine-out NO$_\mathrm{x}^\mathrm{eo}$ emissions by 90\%, while, at the same time, decreasing the H$_2$ consumption by over 16\%, compared to the base H$_2$-vehicle. These significant emission reductions are possible without having to modify the exhaust-gas aftertreatment system, or the optimization of any of the individual drivetrain components, but solely by setting the EMS calibration accordingly.\\
\\
Further research could focus on the development of an online-capable control algorithm for the energy management of the investigated mixed H$_2$-HEV. The goal would be to reach close-to-optimal H$_2$ consumption, while satisfying a NO$_\mathrm{x}^\mathrm{eo}$ emission target. The technique of model predictive control could be used to calculate the optimal torque split in real-time. Another direction of research could be the development and inclusion of an additional dynamic NO$_\mathrm{x}^\mathrm{eo}$ emissions model and the analysis of its influence on the H$_2$-NO$_\mathrm{x}^\mathrm{eo}$ Pareto front of the mixed H$_2$-HEV. Also, the reduced NO$_\mathrm{x}^\mathrm{eo}$ emissions lead to changed requirements for the exhaust-gas aftertreatment system, creating an avenue for exploring the optimization of EMS calibration in conjunction with the design and operation of the exhaust gas aftertreatment system.

\section*{Acknowledgment}
We thank Robert Bosch GmbH for supporting this project.

\bibliographystyle{elsarticle-num-names}
\bibliography{Mixed_HEV_bib}

\end{document}